\documentclass[%
 reprint,preprintnumbers,
 amsmath,amssymb,
 aps,nofootinbib,
 prd,superscriptaddress
 ]{revtex4-2}

\usepackage{hyperref}

\usepackage[utf8]{inputenc}
\usepackage{comment}

\usepackage{graphicx}%
\usepackage{dcolumn}%
\usepackage{bm}%
\usepackage{multibib}
\usepackage{placeins}
\usepackage{xcolor}
\usepackage{orcidlink}
\usepackage{csquotes} %

\usepackage{siunitx} %
\usepackage{cleveref} %
\crefname{figure}{Figure}{Figures}
\crefname{table}{Table}{Tables}
\crefname{section}{section}{sections} %
\crefname{appendix}{appendix}{appendices}
\crefname{equation}{Eq.}{Eqs.}
\Crefname{figure}{Figure}{Figures}
\Crefname{table}{Table}{Tables}
\Crefname{section}{Section}{Sections} %
\Crefname{appendix}{Appendix}{Appendices}
\Crefname{equation}{Eq.}{Eqs.}
\AddToHook{cmd/appendix/before}{\crefalias{section}{appendix}}

\newcommand{\ket}[1]{\ensuremath{| {#1} \rangle }}
\newcommand{\bra}[1]{\ensuremath{\langle {#1} |}}

\newcommand{\GeV}{\giga\electronvolt}

\newcommand{\fm}{\femto\meter}
\newcommand{\as}{\alpha_s}
\newcommand{\LamQCD}{\Lambda_\mathrm{QCD}}

\renewcommand{\vec}[1]{\bm{#1}}

\newcites{SM}{Supplementary Material Bibliography}

\begin{document}

\setlength\abovedisplayskip{10pt}
\setlength\belowdisplayskip{10pt}

\setlength{\parskip}{14pt}
\setlength{\parindent}{0pt}

\title{Inclusive \texorpdfstring{$\bar B_s\mapsto X_{\bar sc} \ell \bar \nu$}{Bs->Xc l nu} decays from lattice QCD:\\ computational strategy and a first physical result}

\author{Alessandro~\surname{De~Santis}~\orcidlink{0000-0002-2674-4222}}
\affiliation{Helmholtz-Institut Mainz, Johannes Gutenberg-Universit{\"a}t Mainz, 55099 Mainz, Germany}
\affiliation{GSI Helmholtz Centre for Heavy Ion Research, 64291 Darmstadt, Germany}

\author{Antonio~\surname{Evangelista}~\orcidlink{0000-0002-3320-3176}}
\affiliation{Department of Physics, University of Cyprus, P.O. Box 20537, 1678 Nicosia, Cyprus}

\author{Gael~\surname{Finauri}~\orcidlink{0009-0005-3859-6991}}
\affiliation{Dipartimento di Fisica, Universit\`a di Torino \& INFN, Sezione di Torino, Via Pietro Giuria 1, I-10125 Turin, Italy}

\author{Roberto~\surname{Frezzotti}~\orcidlink{0000-0001-5746-0065}}
\affiliation{Dipartimento di Fisica \& INFN, Universit\`a di Roma ``Tor Vergata'', Via della Ricerca Scientifica 1, I-00133 Rome, Italy}

\author{Giuseppe~\surname{Gagliardi}~\orcidlink{0000-0002-4572-864X}}
\affiliation{INFN, Sezione di Roma Tre, Via della Vasca Navale 84, I-00146 Rome, Italy}

\author{Paolo~\surname{Gambino}~\orcidlink{0000-0002-7433-4914}}
\affiliation{Dipartimento di Fisica, Universit\`a di Torino \& INFN, Sezione di Torino, Via Pietro Giuria 1, I-10125 Turin, Italy}

\author{Marco~\surname{Garofalo}~\orcidlink{0000-0002-4508-6421}}
\affiliation{HISKP (Theory) \& Bethe Centre for Theoretical Physics, Rheinische Friedrich-Wilhelms-Universit\"at Bonn, Nussallee 14-16, D-53115 Bonn, Germany}

\author{Christiane~Franziska~\surname{Gro\texorpdfstring{\ss}}~\orcidlink{0009-0009-5876-1455}}
\affiliation{HISKP (Theory) \& Bethe Centre for Theoretical Physics, Rheinische Friedrich-Wilhelms-Universit\"at Bonn, Nussallee 14-16, D-53115 Bonn, Germany}

\author{Bartosz~\surname{Kostrzewa}~\orcidlink{0000-0003-4434-6022}}
\affiliation{HISKP (Theory) \& Bethe Centre for Theoretical Physics, Rheinische Friedrich-Wilhelms-Universit\"at Bonn, Nussallee 14-16, D-53115 Bonn, Germany}

\author{Vittorio~\surname{Lubicz}~\orcidlink{0000-0002-4565-9680}}
\affiliation{Dipartimento di Matematica e Fisica, Universit\`a Roma Tre, Via della Vasca Navale 84, I-00146 Rome, Italy}
\affiliation{INFN, Sezione di Roma Tre, Via della Vasca Navale 84, I-00146 Rome, Italy}

\author{Lorenzo~\surname{Maio}~\orcidlink{0000-0001-5267-4064}}
\affiliation{Dipartimento di Fisica \& INFN, Universit\`a di Roma ``Tor Vergata'', Via della Ricerca Scientifica 1, I-00133 Rome, Italy}

\author{Francesca~\surname{Margari}~\orcidlink{0000-0003-2155-7679}}
\affiliation{Dipartimento di Fisica \& INFN, Universit\`a di Roma ``Tor Vergata'', Via della Ricerca Scientifica 1, I-00133 Rome, Italy}

\author{Marco~\surname{Panero}~\orcidlink{0000-0001-9477-3749}}
\affiliation{Dipartimento di Fisica, Universit\`a di Torino \& INFN, Sezione di Torino, Via Pietro Giuria 1, I-10125 Turin, Italy}

\author{Francesco~\surname{Sanfilippo}~\orcidlink{0000-0002-1333-745X}}
\affiliation{INFN, Sezione di Roma Tre, Via della Vasca Navale 84, I-00146 Rome, Italy}

\author{Silvano~\surname{Simula}~\orcidlink{0000-0002-5533-6746}}
\affiliation{INFN, Sezione di Roma Tre, Via della Vasca Navale 84, I-00146 Rome, Italy}

\author{Antonio~\surname{Smecca}~\orcidlink{0000-0002-8887-5826}}
\affiliation{INFN, Sezione di Roma Tre, Via della Vasca Navale 84, I-00146 Rome, Italy}

\author{Javier~\surname{Suarez Sucunza}~\orcidlink{0009-0008-1268-5496}}
\affiliation{HISKP (Theory) \& Bethe Centre for Theoretical Physics, Rheinische Friedrich-Wilhelms-Universit\"at Bonn, Nussallee 14-16, D-53115 Bonn, Germany}

\author{Nazario~\surname{Tantalo}~\orcidlink{0000-0001-5571-7971}}
\affiliation{Dipartimento di Fisica \& INFN, Universit\`a di Roma ``Tor Vergata'', Via della Ricerca Scientifica 1, I-00133 Rome, Italy}

\author{Carsten~\surname{Urbach}~\orcidlink{0000-0003-1412-7582}}
\affiliation{HISKP (Theory) \& Bethe Centre for Theoretical Physics, Rheinische Friedrich-Wilhelms-Universit\"at Bonn, Nussallee 14-16, D-53115 Bonn, Germany}

\begin{abstract}
\centerline{\includegraphics[height=4.2cm]{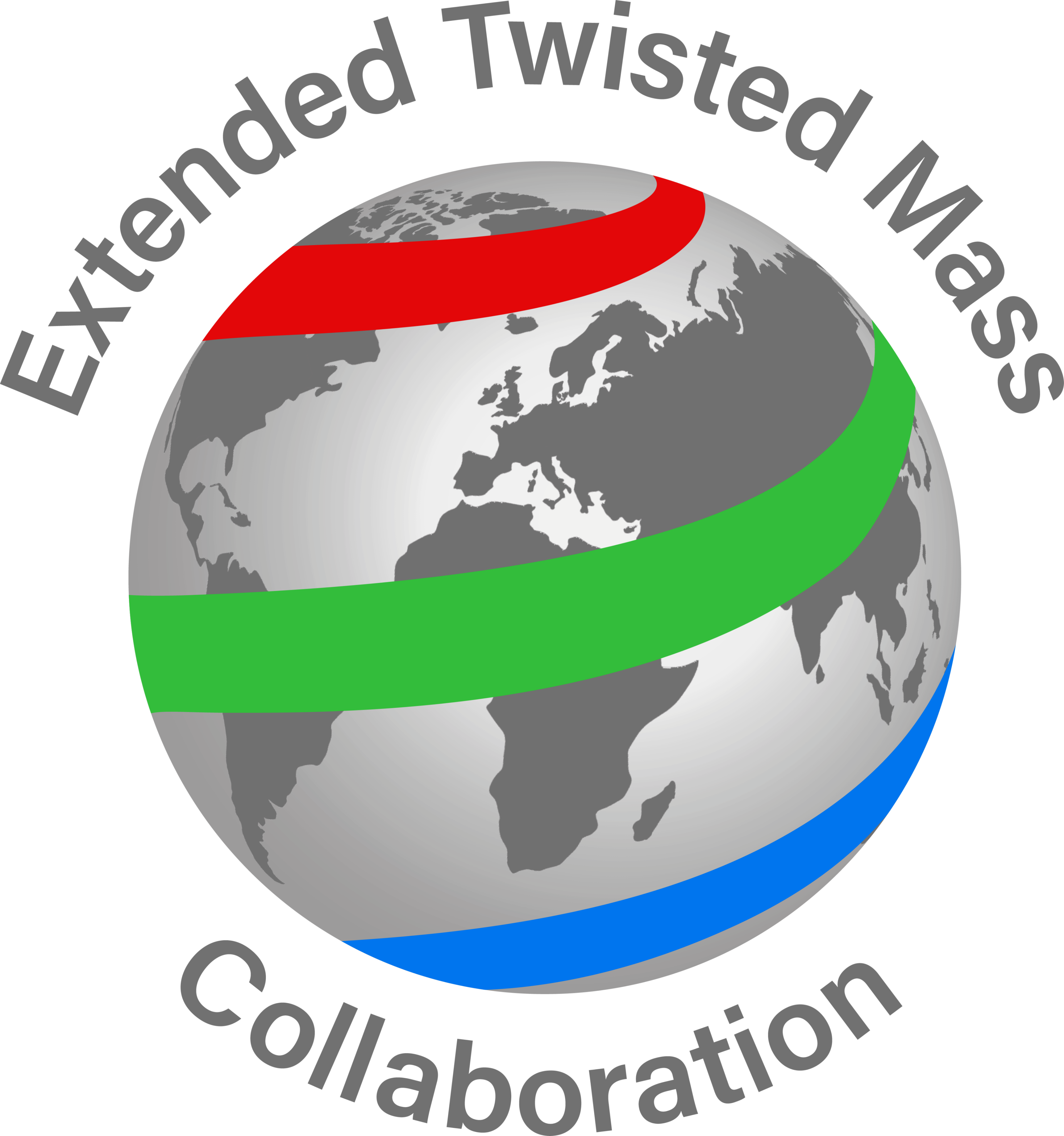}}
\vspace{0.22cm}
We present a strategy to compute the inclusive decay rate for the process $\bar B_s \mapsto X_{\bar sc} \ell \bar{\nu}$ from first principles in lattice QCD. The physical decay rate is obtained from the interpolation of non-perturbative lattice data, obtained at lighter than physical heavy meson masses ($M_{\bar B_s}^\mathrm{max}=4.3$~GeV), with the Operator Product Expansion predictions, which become exact in the limit of infinitely heavy quarks. We also present a new method for the computation of the required lattice four-point correlators, which represents a considerable improvement over the state-of-the-art on the subject. We show the effectiveness of the strategy by performing the calculation on a subset of the available $n_f=2+1+1$ physical-point Extended Twisted Mass Collaboration (ETMC) gauge ensembles. Our current determination of the inclusive decay rate has a 7\% total error, that is dominated by uncertainties due to the relatively limited configuration ensembles considered herein, and can be significantly reduced in the near future.
\end{abstract}

\maketitle

\section{
\label{sec:introduction}
Introduction
}

The Cabibbo--Kobayashi--Maskawa (CKM) matrix element $V_{cb}$ is a fundamental parameter of the Standard Model (SM) and plays a central role in flavour physics. A precise determination of $|V_{cb}|$ is essential not only for describing weak $b\to c$ transitions, but also for performing stringent tests of the SM through flavour-changing neutral-current processes in the $K$- and $B$-meson sectors. Indeed, the uncertainty on $|V_{cb}|$ is currently one of the dominant parametric uncertainties affecting several key flavour observables, such as $\varepsilon_K$, quantities related to $B \mapsto \ell^+\ell^-$ and $K_L\to\pi^0\nu\bar\nu$ decays, etc., thereby limiting their sensitivity to possible new-physics effects~\cite{Buras:2026hbe}. Among the processes that provide the most precise direct determinations of $|V_{cb}|$, semileptonic decays of $B_{(s)}$ mesons into charmed hadronic final states play a pivotal role. Both experimentally and theoretically, the study of semileptonic decays in which a single charmed hadronic channel is selected (exclusive) is very different from that of processes in which the hadronic channels are not separated (inclusive). The two classes of processes provide independent determinations of $|V_{cb}|$, and the long-standing significant tension between them~\cite{Gambino:2020jvv,HeavyFlavorAveragingGroupHFLAV:2024ctg} motivates the considerable effort of the flavour-physics community to reduce theoretical and experimental uncertainties and clarify its origin. This work is part of this collective effort and is focused on the inclusive semileptonic decay $\bar B_s\to X_{\bar s c}\ell\bar\nu$ of the $\bar B_s$ meson into a generic charmed-strange hadronic state $X_{\bar s c}$ and light leptons.

Our final goal is to perform a first-principle lattice QCD calculation of the inclusive decay rate at a phenomenologically relevant level of accuracy. This is a challenging goal, but one within our reach. It is challenging because the theory for the non-perturbative study of inclusive decays on the lattice (which for many years were thought to be inaccessible) has only recently\footnote{The key ingredients were already present in the more general, mathematically-oriented and forward-looking Ref.~\cite{Barata:1990rn}. See also Ref.~\cite{Patella:2024cto} for a recent generalization.} been developed~\cite{Hansen:2017mnd,Hashimoto:2017wqo,Hansen:2019idp,Gambino:2020crt,Gambino:2022dvu}. Moreover, the numerical techniques necessary for the application of this theory require a careful analysis of all error sources, both statistical and systematic. Nevertheless, the results we have obtained in Refs.~\cite{DeSantis:2025yfm,DeSantis:2025qbb} (see also Ref.~\cite{Kellermann:2026sgp}), and the ones that we present in this work, allow us to confidently assert that the goal can be reached.

In Refs.~\cite{DeSantis:2025yfm,DeSantis:2025qbb} we performed the first complete lattice QCD calculation of the inclusive $D_s\mapsto X\ell \bar \nu$ decay rate and of the first two lepton-energy moments. These works demonstrated that it is nowadays possible to study inclusive semileptonic decays on the lattice at the level of accuracy that is required to have an impact on phenomenology. Building on this solid cornerstone, we began addressing the calculation of the $\bar B_s\mapsto X_{\bar s c} \ell \bar \nu$ decay rate. On the one hand, the lattice study of the inclusive $B_{(s)}$ decays requires performing the same steps that we already did in the case of the lighter $D_s$ mesons: (i) the calculation of a set of two- and four-point lattice Euclidean correlators with insertions of interpolating operators and hadronic weak currents; (ii) the extraction of the energy-smeared decay rate from these correlators through a spectral reconstruction technique; (iii) a careful analysis of the systematic errors associated with the spectral reconstruction and with the infinite-volume limit; (iv) a thorough study of the continuum limit and of the associated systematic errors. In a typical lattice setup, it is the last step (the continuum extrapolation) that makes the study of $B_{(s)}$ decays more challenging than that of the $D_{(s)}$ ones.

To calculate the $B_{(s)}$ inclusive decay rate, as in any lattice calculation of $b$-physics hadronic observables, one has to address the two-scale problem associated with the fact that it is hard to simultaneously satisfy the conditions $am_b\ll 1$ (where $a$ is the lattice spacing and $m_b$ the $b$-quark mass) and $\LamQCD L\gg 1$ (where $L^3$ is the lattice spatial volume), which are needed to keep both cutoff and finite-volume effects (FVE) under control. We have decided to address the problem by starting from the state-of-the-art lattice QCD gauge ensembles produced by the Extended Twisted Mass Collaboration (ETMC) with dynamical light, strange and charm quarks with physical masses. These are the ensembles that we already employed in Refs.~\cite{DeSantis:2025yfm,DeSantis:2025qbb} and that allowed us to safely take under control and carefully estimate FVE on the $D_s\mapsto X\ell \bar \nu$ decay rate.
An important result that we present in this work is a numerical strategy for the calculation of the $\bar B_s\mapsto X_{\bar s c} \ell \bar \nu$ decay rate which, starting from these large-volume QCD gauge ensembles, allows to tame cutoff effects and to properly estimate the associated systematic errors.

The Twisted-Mass lattice discretization of the fermionic action that we employ in our simulations leads to automatically $O(a)$ improved physical observables and, consequently, the leading cutoff effects with a physical $b$-quark  would be proportional to $(am_b)^2$. The finest lattice spacing currently available to ETMC is $a\simeq 0.05$~fm. Even at this lattice spacing one finds $(am_b)^2\simeq 1.3$, precluding a direct simulation at the physical $b$-quark mass. Many approaches are available on the market to cope with the lattice ``$b$-problem'' (see the discussion in Ref.~\cite{FlavourLatticeAveragingGroupFLAG:2024oxs} concerning this point) and we are not the only group that is facing the challenge of the lattice calculation of the $\bar B_s\mapsto X_{\bar s c} \ell \bar \nu$ decay rate. In Ref.~\cite{Barone:2023tbl} preliminary results have been obtained by employing the so-called relativistic-heavy-quark (RHQ) action~\cite{El-Khadra:1996wdx,Christ:2006us,Lin:2006ur}.
Our strategy is complementary to the one proposed in Ref.~\cite{Barone:2023tbl} and is built on the analytical results obtained by relying on the Operator Product Expansion (OPE), which become \emph{exact} in the $m_b\mapsto \infty$ limit and that we use to \emph{interpolate} our lattice results obtained at lighter-than-physical $b$-quark masses to the physical point.
Indeed, in the absence of first-principles approaches, OPE techniques~\cite{Manohar:1993qn,Blok:1993va,Bigi:1992su,Bigi:1993fe,Chay:1990da}, that are particularly well motivated in the case of $B_{(s)}$ inclusive decays, have long been the only viable theoretical approach to heavy meson inclusive semileptonic decays. However, we stress that here the OPE formalism is not used to predict the decay rate (by starting from phenomenological inputs).
The OPE analytical results are only employed to guide our numerical interpolation to the physical $\bar B_s$ mass.

In implementing this strategy we encountered, and therefore we had to solve, a rather complicated numerical problem. We observed a rapid deterioration of the statistical signal-to-noise ratio of the required four-point correlators w.r.t.\ the $D_s\mapsto X\ell \bar \nu$ case by increasing the heavy quark mass. After a detailed investigation of the issue, we solved the problem by performing simulations at lighter than physical values of the heavy quark masses, by keeping the $M_{D_s}/M_{\bar{B}_s}$ ratio fixed at its physical value. In this way we managed to keep the phase space volume in dimensionless units constant and, at the same time, to approach the static limit from the direction in which, thanks to asymptotic freedom, the partonic perturbative result becomes exact, without the need of any non-perturbative input, at infinite $m_b$. Moreover, we changed the numerical strategy for the calculation of the required four-point functions w.r.t.\ the previous works on the subject~\cite{Gambino:2020crt,Gambino:2022dvu,Barone:2023tbl,DeSantis:2025yfm,DeSantis:2025qbb,Kellermann:2026sgp}. The new strategy is definitely more expensive from the computational perspective but allows us to completely overcome the signal-to-noise problem. This is the main algorithmic/numerical result that we present in this paper.

Before passing to a detailed illustration of our results, we want to stress that our computational strategy is complementary to the one proposed in Ref.~\cite{Barone:2023tbl} also on the side of the spectral reconstruction algorithm. While Ref.~\cite{Barone:2023tbl} is built on the so-called Chebyshev-polynomials spectral reconstruction technique~\cite{Barata:1990rn,Bailas:2020qmv}, we adopt, as already done in Refs.~\cite{DeSantis:2025yfm,DeSantis:2025qbb} (see also~\cite{ExtendedTwistedMassCollaborationETMC:2022sta,Evangelista:2023fmt,Bonanno:2023ljc,Bonanno:2023thi,ExtendedTwistedMass:2024myu}), the so-called Hansen--Lupo--Tantalo (HLT) algorithm of Ref.~\cite{Hansen:2019idp}. From the theoretical perspective the two approaches are closely related. On the algorithmic side, however, the procedures used to estimate the systematic uncertainties associated with the spectral reconstruction differ substantially. This, together with the fact that we rely on a different numerical strategy to solve the $b$-problem, is an important added value. Indeed, when both projects will be fully completed, there will be two totally independent (because of the different lattice discretizations and analysis techniques) first-principles results for the $\bar B_s\mapsto X_{\bar s c} \ell \bar \nu$ decay rate.

In this paper, we present a result for the decay rate that is not yet final: it does not include the systematic uncertainties associated with finite-volume effects and has been obtained using only three lattice spacings. Nevertheless, we have already carried out a careful analysis of the different stages of the calculation, together with a conservative estimate of the total uncertainty. Thus, despite the limitations implied by the resulting 7\% error --which we expect to reduce in the near future-- our current results can already be used reliably in phenomenological applications.

The rest of the paper is organized as follows. In \cref{sec:inclusivenotation} we set our notation and recall the formulae for the extraction of the decay rate from lattice correlators. In \cref{sec:OPE} we recall the OPE analytical formulae that we will then use to interpolate our lattice data at the physical point. In \cref{sec:latticecorrelators} we introduce and explain the new numerical strategy for the computation of the required lattice correlators. In \cref{sec:resultswithhlt,sec:results} we discuss the analysis of the lattice data which we then use in \cref{sec:interpolation} to interpolate the physical result. We finally draw our conclusions and present our outlooks in \cref{sec:conclusion}.

\section{
\label{sec:inclusivenotation}
Inclusive semileptonic decays of heavy mesons
}

In order to obtain a first-principles theoretical prediction for the physical $\bar B_s\mapsto X_{\bar s c} \ell \bar \nu$ decay rate, we perform lattice simulations at lighter than physical values of the bottom and charm quark masses. We then interpolate our numerical results to the physical point by relying on the OPE analytical predictions. These, thanks to asymptotic freedom, are exact (independent from non-perturbative inputs) in the limit in which the bottom and charm quark masses are both sent to infinity keeping their ratio constant.

On each gauge ensemble, and therefore at fixed lattice spacing $a$ and fixed physical volume (see \cref{tab:iso_EDI_FLAG}), we consider a sequence $(m_i,m_f)$ of pairs of unphysical valence heavy quark masses such that $am_{f}<am_i\ll 1$. For each pair $(m_i,m_f)$ we compute the masses $(M_I,M_F)$ of the decaying pseudoscalar meson $I$, with valence flavour quantum numbers $\bar s i$, and of the pseudoscalar meson $F$, with valence flavour quantum numbers $\bar s f$, which is the lightest final hadronic state contributing to the inclusive process $I\mapsto X_f \ell \bar \nu$. In this way we establish the non-perturbative correspondence
\begin{flalign}
(m_i,m_f)
\, \longleftrightarrow \, (M_I,M_F)\;.
\end{flalign}
The physical point is the one in which $(m_i,m_f)^\mathrm{phys}=(m_b,m_c)$ and, correspondingly, $(M_I,M_F)^\mathrm{phys}=(M_{\bar B_s},M_{D_s})$, where $M_{\bar B_s}$ is the physical $\bar B_s$ mass and $M_{D_s}$ is the physical $D_s$ mass.

For each simulated pair $(m_i,m_f)$ we then compute the decay rate $\Gamma=\Gamma[I\mapsto X_{f} \ell \bar \nu]$ by using the formulae presented and described in full details in Ref.~\cite{DeSantis:2025qbb}. Here, in order to make this article self-contained, and to establish the mapping between the notations used in the two works, we report the essential formulae and refer to Ref.~\cite{DeSantis:2025qbb} for further details.

\begin{figure}[t!]
    \centering
    \includegraphics[width=0.7\linewidth]{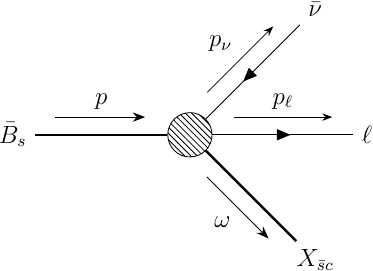}
    \caption{Kinematics of the $\bar{B}_s$ inclusive semileptonic decay studied in this work.}
    \label{fig:kinematic}
\end{figure}

We work in the rest-frame of the decaying $I$ meson and call
\begin{flalign}
&
p=M_I(1,\vec 0)\;,
\qquad\quad
\omega=M_I(\omega_0,\vec \omega)\;,
\nonumber \\[8pt]
&
p_\ell=M_I(e_\ell,\vec k_\ell)\;,
\qquad
p_\nu=M_I(e_\nu,\vec k_\nu)\;,
\end{flalign}
the four-momenta of the initial state $I$, of the generic hadronic state $X_f$, of the lepton and of the neutrino, so that the energy-momentum conservation relation $p=p_\ell+p_\nu+\omega$ (see \cref{fig:kinematic}) implies
\begin{flalign}
\omega_0=1-e_\ell-e_\nu\;,
\qquad
\vec \omega=-\vec k_\ell -\vec k_\nu\;.
\end{flalign}
We work in the approximation in which the charged lepton is massless and therefore we have $\vec k_\ell^2=e_\ell^2$ as well as $\vec k_\nu^2=e_\nu^2$.

\begin{figure}
    \centering
    \includegraphics[width=\linewidth]{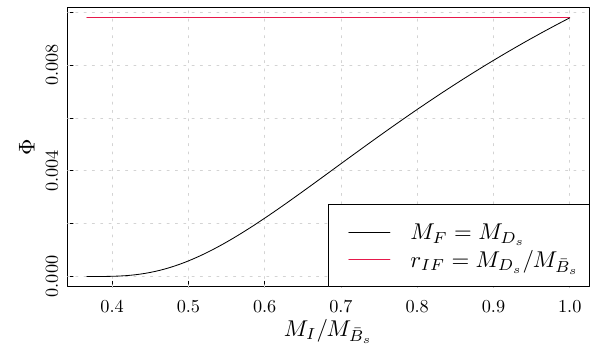}
    \caption{In our simulation strategy we vary both the initial and final meson masses $M_I$ and $M_F$ by tuning the $(m_i,m_f)$ quark masses so that the condition $M_I/M_F=M_{D_s}/M_{\bar B_s}$ is satisfied. In this strategy the volume of the phase space available for the decay remains constant (red line). The black curve corresponds to the strategy (not adopted in this work) in which only $M_I$ is varied while $M_F$ is kept constant at its physical value, i.e.\ $M_F=M_{D_s}$.}
    \label{fig:availablephasespace}
\end{figure}

The total decay rate can be expressed as
\begin{flalign}
&
\Gamma=
\int_0^{(\vert\vec \omega\vert^\mathrm{max})^2} d\vec \omega^2
\int_{\omega^\mathrm{min}}^{\omega^\mathrm{max}} d\omega_0
\int_{e_\ell^\mathrm{min}}^{e_\ell^\mathrm{max}} de_\ell
\frac{d\Gamma}{d\vec \omega^2 d\omega_0 de_\ell}\;,
\label{eq:gammfgtripleint}
\end{flalign}
and we are now going to provide, in turn, the explicit formulae for the phase-space integration, for the triple-differential decay rate and for the extraction of the total rate from Euclidean correlators.

\subsection{
\label{sec:phasespace}
The phase-space
integration
}

The phase-space integration limits to be used in \cref{eq:gammfgtripleint} are given by the following expressions
\begin{flalign}
&
e_\ell^\mathrm{min}=\frac{1-\omega_0-\sqrt{\vec \omega^2}}{2}\;,
\qquad
e_\ell^\mathrm{max}=\frac{1-\omega_0+\sqrt{\vec \omega^2}}{2}\;,
\nonumber \\[8pt]
&
\omega^\mathrm{min}=\sqrt{r_{IF}^2+\vec \omega^2}-0^+\;,
\qquad
\omega^\mathrm{max}=1-\sqrt{\vec \omega^2}\;,
\nonumber \\[8pt]
&
\vert\vec \omega\vert^\mathrm{max}= \left(\frac{1-r_{IF}^2}{2}\right)+0^+\;,
\label{eq:limits1}
\end{flalign}
where
\begin{flalign}
r_{IF}=\frac{M_F}{M_I}\;.
\end{flalign}
We refer to Ref.~\cite{DeSantis:2025qbb} for a detailed explanation of the theoretical motivation behind the $0^+$-prescription made explicit in \cref{eq:limits1}.

The strategy of our lattice calculation has been built on the important observation that the volume of the phase-space available for the decay,
\begin{flalign}
&
\Phi=
\int_0^{(\vert\vec \omega\vert^\mathrm{max})^2} d\vec \omega^2
\int_{\omega^\mathrm{min}}^{\omega^\mathrm{max}} d\omega_0
\int_{e_\ell^\mathrm{min}}^{e_\ell^\mathrm{max}} de_\ell
\nonumber \\[8pt]
&
=
\frac{
(1-r_{IF}^2)
(1-5r_{IF}^2-2r_{IF}^4)
-6r_{IF}^4\log(r_{IF}^2)
}
{48}\;,
\label{eq:phasespacevolume}
\end{flalign}
is entirely determined by $r_{IF}$ and (obviously) vanishes at $r_{IF}=1$.

On all of our $n_f=2+1+1$ gauge ensembles, the condition $am_c\ll 1$ is satisfied (see \cref{tab:iso_EDI_FLAG}). Therefore, in choosing the sequence of quark masses to be simulated, we had the option of keeping the final quark mass fixed at its physical value, i.e.\ of setting $m_f=m_c$ and increasing only the initial quark mass. In this strategy, though, the volume of the phase-space available for the decay is very small when $m_i$ is close to $m_f=m_c$ and, consequently, $r_{IF}\simeq 1$ (see \cref{{eq:phasespacevolume}} and \cref{{fig:availablephasespace}}).

In order to avoid coping with this problem we have instead chosen to keep the phase-space volume $\Phi$ fixed by simulating a sequence of increasingly heavy initial quark masses and by determining the corresponding final quark masses by imposing the condition
\begin{flalign}
r_{IF}=r_{IF}^\mathrm{phys}\equiv\frac{M_{D_s}}{M_{\bar B_s}} = 0.36676(1)
\quad \mapsto \quad m_f=m_f(m_i)\;.
\label{eq:rIFcondition}
\end{flalign}
We provide all the details of the numerical procedure that we use to impose this condition in \cref{{sec:masstuning}}. Here we stress that this strategy is particularly advantageous from the OPE perspective. Indeed, by sending both $m_i$ and $m_f$ to infinity at fixed $r_{IF}$ the decay rate can be \emph{exactly} calculated at tree-level in QCD perturbation theory and, by relying on this observation, we can turn the extrapolation of our numerical results to the physical point into an interpolation.

\subsection{The differential decay rate}
The differential decay rate
\begin{flalign}
&
\frac{d\Gamma}{d\vec \omega^2 d\omega_0 de_\ell}
=
\frac{M_I^5 G_F^2 S_\mathrm{EW}}{32\pi^4}\Bigg(
\nonumber \\[8pt]
&
-
2\left\{(1-\omega_0)^2-\vec \omega^2 \right\} h^{(1)}
+
\left\{
\vec \omega^2-(1-\omega_0-2e_\ell)^2
\right\} h^{(2)}
\nonumber \\[8pt]
&
+
2\left[(1-\omega_0)^2-\vec \omega^2 \right]\left[2e_\ell-(1-\omega_0)\right] h^{(5)}
\Bigg)\;,
\label{eq:tripleGhs}
\end{flalign}
which we define without including the CKM matrix element,
depends upon\footnote{$G_F$ is the Fermi constant and $S_\mathrm{EW}=1.023$ is the factor that accounts for the $O(\alpha_\mathrm{em})$ corrections to the total width~\cite{Bigi:2023cbv}. It includes the electroweak corrections to the Wilson coefficient, Sirlin's logarithm~\cite{Sirlin:1981ie}, and the perturbative QED corrections computed in Ref.~\cite{Bigi:2023cbv}.} the hadronic form-factors which parametrize the so-called hadronic tensor
\begin{flalign}
&
H_{\mu\nu}(p,\omega)
=
\frac{(2\pi)^4}{2M_I}
\bra{I(p)}J_\mu^\dagger(0)\, \delta^4(\mathbb{P}-\omega)\, J^\nu(0)\ket{I(p)}
\;,
\label{eq:hmunudef}
\end{flalign}
where $\mathbb{P}=(H,\vec P)$ is the QCD (including the unphysical pair of valence quarks) four-momentum operator and
\begin{flalign}
&
J^\mu(x)
=\bar \psi_f(x) \gamma^\mu(1-\gamma^5) \psi_i(x)
\end{flalign}
is the hadronic weak current. Indeed, by relying on Lorentz and time-reversal covariance, the hadronic tensor $H^{\mu\nu}(p,\omega)$ can be decomposed into five dimensionless form-factors, according to
\begin{flalign}
&
M_I^3 H^{\mu\nu}(p,\omega)
=
g^{\mu\nu} M_I^2 h^{(1)}
+
p^\mu p^\nu h^{(2)}
\nonumber \\[8pt]
&
+
(p-\omega)^\mu (p-\omega)^\nu h^{(3)}
+
\left\{p^\mu (p-\omega)^\nu+(p-\omega)^\mu p^\nu\right\} h^{(4)}
\nonumber \\[8pt]
&
+
i\epsilon^{\mu\nu\alpha\beta} p_\alpha (p-\omega)_\beta h^{(5)}\;,
\label{eq:hff}
\end{flalign}
where $\epsilon^{\alpha\beta\gamma\delta}$ is the totally antisymmetric four-index Levi--Civita symbol, with $\epsilon^{0123}=1$. The form-factors are real and depend on the scalars $p\cdot \omega$ and $\omega^2$ which, in the rest frame of the $I$ meson, are functions of $\omega_0$ and $\vec \omega^2$. We therefore have
\begin{flalign}
&
h^{(i)}\equiv
h^{(i)}
(\omega_0,\vec \omega^2)
\equiv
h^{(i)}
\left(\frac{\LamQCD}{M_I},\frac{M_F}{M_I};\omega_0,\vec \omega^2\right)\;,
\nonumber \\[8pt]
&
i=1,\cdots,5\;,
\label{eq:ffdeps}
\end{flalign}
where in the second equality we used the fact that in our convention the form factors are dimensionless and have explicitly shown the dependence on the meson masses $(M_I,M_F)$, inherited from the dependence of $H^{\mu\nu}(p,\omega)$ upon the QCD scale $\LamQCD$ and all the quark masses (which instead we left implicit in previous formulae).

\subsection{
\label{sec:differentialrate}
Extracting the decay rate from Euclidean correlators
}
On the lattice it is not possible to directly compute the hadronic tensor $H_{\mu\nu}(p,\omega)$ (see \cref{eq:hmunudef}) which, theoretically, is a hadronic spectral density and, mathematically, a tempered distribution. It is instead possible to access its Laplace transform,
\begin{flalign}
&
\hat H_{\mu\nu}(t;p,\vec \omega)
=
\int_{\omega^\mathrm{min}}^\infty d\omega_0\, e^{-\omega_0 (M_I t)}\,
H_{\mu\nu}(p,\omega)
\;,
\label{eq:hmunut}
\end{flalign}
an amputated Euclidean hadronic correlator and a properly smeared distribution which, as such, is a numerically computable quantity. For this reason, in order to perform a first-principles lattice calculation of the decay rate it is first necessary to derive a mathematical representation of $\Gamma$ in terms of suitable linear combinations of the different components of $\hat H_{\mu\nu}(t;p,\vec \omega)$.

To this end, by relying upon the fact that the hadronic form-factors do not depend on the lepton energy, we start by performing analytically the $e_\ell$ integral in \cref{{eq:gammfgtripleint}}. The result is
\begin{flalign}
&
\frac{1}{\bar \Gamma}
\frac{d \Gamma}{d \omega^0 d \vec \omega^2}
=
\nonumber \\[8pt]
&
\vert \vec \omega\vert^3\, Z^{(0)}
+
\vert \vec \omega\vert^2 (\omega^\mathrm{max}-\omega_0)\, Z^{(1)}
+
\vert \vec \omega\vert (\omega^\mathrm{max}-\omega_0)^2\, Z^{(2)},
\label{eq:doubleGhs}
\end{flalign}
where
\begin{flalign}
\bar \Gamma= \frac{M_I^5 G_F^2 S_\mathrm{EW}}{48\pi^4}\;,
\end{flalign}
and where, by considering the two unit vectors $\vec{\hat n}_r$ that are orthonormalized and orthogonal to $\vec{\hat \omega} = \vec \omega/\vert \vec \omega\vert$, i.e.
\begin{flalign}
\vec{\hat n}_r\cdot \vec{\hat n}_s=\delta_{rs}\;,
\qquad
\vec{\hat n}_r\cdot \vec{\hat \omega}=0\;,
\qquad
r,s=1,2\;,
\end{flalign}
we have introduced the following three linear combinations\footnote{In Ref.~\cite{DeSantis:2025qbb} the $Z^{(p)}$'s  have been defined in terms of other five linear combinations of the different components of the hadronic tensor, called $\mathcal{Y}^{(i)}$, that are useful in order to invert the system of \cref{eq:hff} and  to parametrize the so-called lepton-energy moments which we do not compute in this paper. The definition of the $Z^{(p)}$'s given here is more direct but totally equivalent to the one given in Ref.~\cite{DeSantis:2025qbb}.} of the
different components of $H_{\mu\nu}$
\begin{flalign}
&
Z^{(0)}=
M_I\left[
H^{00} +
\sum_{i,j=1}^3 \hat \omega^i \hat \omega^j H^{ij}
+2\sum_{i=1}^3 \hat \omega^i H^{0i}
\right]\;,
\nonumber \\[12pt]
&
Z^{(1)}=
2M_I\left[
\sum_{i,j=1}^3 \hat \omega^i \hat \omega^j H^{ij}
\right.
\nonumber \\[4pt]
&
\qquad\qquad
\left.
+\sum_{r=1}^2\sum_{i,j=1}^3 \hat n^i_r \hat n^j_r H^{ij}
+\sum_{i=1}^3 \hat \omega^i H^{0i}
\right]\;,
\nonumber \\[12pt]
&
Z^{(2)}=M_I\left[
\sum_{i,j=1}^3 \hat \omega^i \hat \omega^j H^{ij}
+\sum_{r=1}^2\sum_{i,j=1}^3 \hat n^i_r \hat n^j_r H^{ij}
\right]\;.
\label{eq:Zgamma}
\end{flalign}
In \cref{eq:doubleGhs,eq:Zgamma}, as already done in the case of the form-factors, see \cref{eq:ffdeps}, we used the compact notation
\begin{flalign}
&
Z^{(p)}
\equiv Z^{(p)}(\omega_0,\vec \omega^2)
\equiv Z^{(p)}
\left(\frac{\LamQCD}{M_I};\frac{M_F}{M_I},\omega_0,\vec \omega^2\right)\;,
\nonumber \\[8pt]
&
p=0,1,2\;.
\label{eq:Zdeps}
\end{flalign}
On the lattice the distributions $Z^{(p)}(\omega_0,\vec \omega^2)$ appear as the hadronic spectral densities of the amputated Euclidean correlators
\begin{flalign}
&
\hat Z^{(p)}(t;\vec \omega)
=
\int_{\omega^\mathrm{min}}^\infty d\omega_0\, e^{-\omega_0 (M_I t)}\,
Z^{(p)}(\omega_0,\vec \omega^2)
\;,
\label{eq:Zpt}
\end{flalign}
which are obtained by relying on the linearity of the Laplace transform, see \cref{eq:hmunut}, and by plugging $\hat H_{\mu\nu}(t;p,\vec \omega)$ instead of $H_{\mu\nu}(p,\vec \omega)$ in  \cref{eq:Zgamma}.

In order to get the representation of $\Gamma$ in terms of Euclidean correlators we are looking for, we now introduce the kernels
\begin{flalign}
\Theta_\sigma^{(p)}(x) = x^p\, \Theta_\sigma(x)\;,
\label{eq:defthetap}
\end{flalign}
where $\Theta_\sigma(x)$ is any Schwartz\footnote{That is, infinitely differentiable and vanishing, together with all of its derivatives, faster than any power of $x$ in the limit $x\mapsto -\infty$.} representation of the Heaviside step-function $\theta(x)$, which depends smoothly on the smearing parameter $\sigma$ and which is such that
\begin{flalign}\label{eq:heaviside}
\lim_{\sigma\mapsto 0} \Theta_\sigma(x)=\theta(x)\;.
\end{flalign}
The introduction of this mathematical device allows one to trade the $\omega_0$ phase-space integral, to be performed in the compact interval $[\omega^\mathrm{min},\omega^\mathrm{max}]$ (see \cref{eq:gammfgtripleint,eq:limits1}), for convolutions of the distributions $Z^{(p)}(\omega_0,\vec \omega^2)$ with smooth smearing kernels,
\begin{flalign}
&
\frac{1}{\bar \Gamma}
\frac{d \Gamma^{(p)}(\sigma)}{d \vec \omega^2}
=
\nonumber \\[8pt]
&
\vert \vec \omega \vert^{3-p}\,
\int_{\omega^\mathrm{min}}^\infty d\omega_0\,
\Theta_\sigma^{(p)}(\omega^\mathrm{max}-\omega_0)\, Z^{(p)}(\omega_0,\vec \omega^2),
\label{eq:dGZint}
\end{flalign}
and with a limiting procedure,
\begin{flalign}
\Gamma
=
\int_0^{(\vert\vec \omega\vert^\mathrm{max})^2} d\vec \omega^2\,
\sum_{p=0}^2
\lim_{\sigma\mapsto 0}\left\{
\frac{d\Gamma^{(p)}(\sigma)}{d \vec \omega^2}
\right\}\;.
\label{eq:gammalimit}
\end{flalign}
Moreover, the Stone--Weierstrass theorem guarantees that for any positive value of the length scale $a$ the smooth kernels $\Theta_\sigma^{(p)}(\omega^\mathrm{max}-\omega_0)$ can be exactly represented according to
\begin{flalign}
\Theta_\sigma^{(p)}(\omega^\mathrm{max}-\omega_0)
=
\lim_{N\mapsto \infty}\sum_{n=1}^{N} g^{(p)}_n(N)\, e^{-\omega_0 (a M_I) n} \;.
\label{eq:stoneZ}
\end{flalign}
The coefficients $g^{(p)}_n(N)$ appearing in the previous formula have to be read as the coordinates of the kernels $\Theta_\sigma^{(p)}(\omega^\mathrm{max}-\omega_0)$ on the discrete set of basis-functions $\exp[-\omega_0 (a M_I) n]$. By identifying $a$ with the lattice spacing\footnote{See Ref.~\cite{Patella:2024cto} for the generalization of this strategy to the case in which the length scale $a$, called $\tau$ in that paper, is kept constant in physical units. It is important to stress that when $a$ is the lattice spacing the $N\mapsto \infty$ limit must be performed \emph{before} the continuum limit.} and by considering the discrete Euclidean times $t=an$, the connection between the decay rate and lattice Euclidean correlators can now be established.

From \cref{eq:Zpt,eq:dGZint,eq:stoneZ} we have
\begin{flalign}\label{eq:lim_N_gamma}
\frac{1}{\bar \Gamma}
\frac{d \Gamma^{(p)}(\sigma)}{d \vec \omega^2}
=\vert \vec \omega \vert^{3-p}\,
\lim_{N\mapsto \infty}\sum_{n=1}^{N} g^{(p)}_n(N)\,
\hat Z^{(p)}(a n,\vec \omega^2)\; ,
\end{flalign}
which, combined with \cref{eq:gammalimit}, gives
\begin{flalign}
&
\Gamma
=
\bar \Gamma
\int_0^{(\vert\vec \omega\vert^\mathrm{max})^2} d\vec \omega^2\,
\sum_{p=0}^2 \lim_{\sigma\mapsto 0}\Bigg\{
\nonumber \\[8pt]
&
\qquad
\,
\vert \vec \omega \vert^{3-p}\,
\left[
\lim_{N\mapsto \infty}\sum_{n=1}^{N} g^{(p)}_n(N)\,
\hat Z^{(p)}(a n,\vec \omega^2)
\right]
\Bigg\}
\;.
\label{eq:gammalimit2}
\end{flalign}

The effectiveness of this representation has been demonstrated in Ref.~\cite{DeSantis:2025qbb} and here we use the same numerical strategy:
in order to determine the coefficients $g^{(p)}_n(N)$, and to study numerically the $N\mapsto \infty$ limit at fixed $\sigma>0$ and the associated systematic errors, we use the HLT algorithm of Ref.~\cite{Hansen:2019idp} (see section VII of~\cite{DeSantis:2025qbb}); in order to perform the necessary $\sigma\mapsto 0$ extrapolations we rely on the asymptotic formulae derived and discussed in section V of~\cite{DeSantis:2025qbb}; the $\vec \omega^2$ integral is evaluated numerically by computing the required Euclidean correlators at several values of the momentum $\vec \omega$ of the final hadronic state $X_f$. All these steps are illustrated and discussed in the following sections.

The main difference of the present work w.r.t. Ref.~\cite{DeSantis:2025qbb} is the interpolation required in order to obtain the physical $\Gamma[\bar B_s\mapsto X_{\bar s c} \ell \bar \nu]$ decay rate by starting from the results obtained at lighter than physical $b$-quark masses. This additional step, which we perform by relying on the OPE formulae, is actually the main focus of this paper and is thoroughly discussed in \cref{sec:OPE,sec:interpolation} from both the theoretical and numerical perspectives.

\section{Operator Product Expansion}
\label{sec:OPE}
In this section we discuss the OPE for $\bar{B}_s \mapsto X_{\bar{s}c} \ell \bar{\nu}$ inclusive decays.
This is a theoretical framework in which the decay rate is obtained as a combined double expansion: the perturbative one in $\as$ and a power series in $\LamQCD/m_b$, exploiting the hierarchy $m_b \gg \LamQCD$.
The series in $\LamQCD/m_b$ is characterized by the presence of non-perturbative matrix elements of local operators (of increasing dimension) between $\bar{B}_s$ meson states.
However, at the leading power no non-perturbative information is needed as the theoretical prediction is purely partonic.
This is due to the fact that the lowest dimensional operator in the OPE is a conserved current in QCD
\begin{equation}
    \langle \bar{B}_s(p) |\bar{b} \gamma^\mu (1-\gamma^5) b | \bar{B}_s(p) \rangle = 2 p^\mu\,.
\end{equation}
This feature allows us to predict with purely perturbative methods the $I \mapsto X_f \ell \bar{\nu}$ decay rate for very heavy meson masses such that $\LamQCD/M_I \mapsto 0$.
The key idea is to pair the lattice points computed at $M_I < M_{\bar{B}_s}$ in a fully non-perturbative manner with the theoretical information in the infinite meson mass limit $M_I \mapsto \infty$.
This turns the extrapolation to the physical point $M_I = M_{\bar{B}_s}$ into an interpolation, drastically improving the systematic uncertainty on the extracted result.

The partonic calculation is expressed in terms of quark masses, while the lattice calculation presented in this work eliminates the dependence on the quark masses in favour of the meson masses, as explained in~\cref{sec:inclusivenotation}.
For heavy quarks, we know from heavy-quark effective theory (HQET) the relation between the on-shell quark mass and the meson mass~\cite{Neubert:1993mb}
\begin{equation}
\label{eq:quarkmesonMass}
    M_I = m_i + \bar{\Lambda} + O(\LamQCD^2/m_i)\,,
\end{equation}
where $\bar{\Lambda}$ is a non-perturbative parameter of order $\LamQCD$ which does not depend on the quark mass.
Hence in the limit $(m_i,M_I) \mapsto \infty$ we can identify the heavy quark mass $m_i$ with the heavy meson mass $M_I$, generating power corrections which vanish linearly with $\bar{\Lambda}/M_I$.

The partonic decay rate is known perturbatively up to
$O(\as^3)$~\cite{Fael:2020tow}.
The renormalization scale $\mu_s$ at which $\as(\mu_s)$ needs to be evaluated
is tied to the physical scale of the process, namely $M_I$.
Therefore, in the limit $M_I \mapsto \infty$, also the perturbative corrections will vanish due to asymptotic freedom of QCD.
However the vanishing of the perturbative corrections is governed by $\as(\mu_s\mapsto \infty) \sim (\log \frac{\mu_s}{\LamQCD})^{-1}$, which is much slower than the vanishing of the power corrections.
For this reason, in~\cref{sec:interpolation} we will include the perturbative information at different orders in $\as$ to test the robustness of the interpolation strategy and to estimate the associated systematic errors.
In the on-shell scheme the partonic decay rate is expressed as
\begin{flalign}
&
    \Gamma_{\rm part} =
\nonumber \\[8pt]
&
    \frac{\pi}{4} \bar{\Gamma}\biggl\{X_0 + \frac{\as(\mu_s)}{\pi} X_1  + \frac{\as^2(\mu_s)}{\pi^2} X_2 + \frac{\as^3(\mu_s)}{\pi^3} X_3 \biggr\}\;,
    \label{eq:partonicres}
\end{flalign}
with
\begin{flalign}
    X_0 = 1 - 8 r_{IF}^2 + 8 r_{IF}^6 - r_{IF}^8 -24 r_{IF}^4 \log r_{IF}\;,
\end{flalign}
while the explicit expressions of the other $X_i$ terms can be found in Refs.~\cite{Nir:1989rm,Pak:2008cp,Dowling:2008mc,Fael:2020tow}.
In the static limit $\mu_s \propto M_I \mapsto \infty$, where all perturbative corrections vanish, we find
\begin{equation}
    \frac{\Gamma_{\rm part}}{\bar{\Gamma}}\biggl|_{M_I \mapsto \infty} = 0.29737(2) \,,
\end{equation}
which will be used as an anchor for reaching the physical point $M_I = M_{\bar{B}_s}$ from the non-perturbative lattice data.

Notice that the perturbative series of \cref{eq:partonicres} in the on-shell scheme has serious convergence issues due to the known renormalon problem of the quark pole mass (see e.g. Ref.~\cite{Beneke:1998ui}).
This is often circumvented by choosing a short-distance renormalization scheme for the heavy quark masses, in particular the kinetic scheme for the $b$-mass~\cite{Bigi:1996si,Fael:2020iea}.
However, the differences between these schemes become irrelevant at high $M_I$. In the static limit, the partonic on-shell result becomes exact and can still be trusted whenever power corrections are negligible. The calculations of \cref{eq:partonicres} in the on-shell and kinetic schemes lead to compatible results for $M_I \ge 10M_{B_s}$. Therefore, we are allowed to use $\Gamma_\mathrm{part}$ in the interpolation procedure carried out in \cref{sec:interpolation}, which we have chosen due to the more natural connection between the pole quark masses and the heavy meson masses,  \cref{eq:quarkmesonMass}.

\section{
\label{sec:latticecorrelators}
Lattice Correlators
}

\begin{table*}[t]
\centering
\begin{tabular}{lcccccccl}
ensemble & $L/a$ & $a~[\rm fm]$ & $L~[\rm fm]$ &  $am_{ud}$ & $am_{s}$ & $am_{c}$ & $am_\mathrm{cr}$ & ~ status ~ \\
\hline \\[-2pt]
B64    & $64$ & ~ $0.07948(11)$ ~  & ~ 5.09 ~ &~ 0.0006669(28)  ~ & ~0.018267(53)~   &  ~0.23134(52) ~ & ~ -0.4138934(46)	~ & ~ completed ~ \\[4pt]
 C80    & $80$    & ~ $0.06819(14)$ ~   & 5.46 &~ 0.0005864(34) ~
   &  ~0.016053(67)~  &  ~0.19849(64)	~  & ~ -0.3964534(41)  ~& ~ completed ~ \\[4pt]
D96    & $96$    & ~ $0.056850(90)$ ~  & 5.46 &~ 0.0004934(24) ~
   &  ~0.013559(39)~  &  ~0.16474(44)	~  & ~ -0.3761252(39) ~ & ~ completed ~ \\[4pt]
 E112    & $112$    & ~ $0.04892(11)$ ~  & 5.48 &~ 0.0004306(23) ~  &  ~0.011787(55)~  &   ~0.14154(54)	~ & ~  -0.3613136(75) ~ & ~ ongoing ~ \\[4pt]
 B48    & $48$ & ~ $0.07948(11)$ ~  & ~ 3.82 ~ &~ 0.0006669(28)  ~ & ~0.018267(53)~   &  ~0.23134(52) ~ & ~ -0.4138934(46)	~ & ~ planned ~ \\
 B96    & $96$    & ~ $0.07948(11)$ ~  & ~ 7.63 ~ & ~ 0.0006669(28)  ~ & ~0.018267(53)~   &  ~0.23134(52) 	~ & ~ -0.4138934(46) ~ & ~ planned ~ \\[8pt]
\hline
\end{tabular}

\caption{
ETMC gauge ensembles used in this work. We give the values of the lattice spacing $a$, of the spatial lattice extent $L$, of the simulated bare light ($m_u=m_d=m_{ud}$), strange ($m_s$) and charm ($m_c$) quark masses and of the critical mass $m_\mathrm{cr}$. The temporal extent of the lattice is always $T=2L$. Details concerning the determination of the lattice spacing and of the quark masses can be found in~\cite{ExtendedTwistedMass:2024nyi}.}
\label{tab:iso_EDI_FLAG}
\end{table*}

The gauge ensembles on which we calculate the Euclidean correlators needed in this project are listed in \cref{tab:iso_EDI_FLAG}. These are large-volume physical-point isospin-symmetric QCD (isoQCD) ensembles that have been generated~\cite{Alexandrou:2018egz,ExtendedTwistedMass:2020tvp, ExtendedTwistedMass:2021qui,Finkenrath:2022eon} by the ETMC with $n_f = 2 + 1 + 1$ flavours of Wilson-Clover Twisted Mass (TM) sea quarks~\cite{Frezzotti:2000nk,Frezzotti:2003xj}. The bare parameters of the simulations have been tuned to match our scheme of choice for defining isoQCD, the so-called Edinburgh/FLAG consensus~\cite{FlavourLatticeAveragingGroupFLAG:2024oxs}, and therefore to match the inputs $M_\pi=135.0$~MeV, $M_K=494.6$~MeV, $M_{D_s}=1967$~MeV and $f_\pi=130.5$~MeV. The $b$-quark is not dynamical in our calculation and, as anticipated in the previous sections and explained in full details below, has been simulated in valence with a lighter than physical mass.

The results that we present in this paper have been obtained on the B64 ($a\simeq 0.08$~fm, $L\simeq 5.1$~fm), C80 ($a\simeq 0.07$~fm, $L\simeq 5.5$~fm) and D96 ($a\simeq 0.06$~fm, $L\simeq 5.5$~fm) ensembles on which we have already completed the evaluation of the necessary correlators (see \cref{tab:iso_EDI_FLAG}). We are currently completing the calculation of the correlators on the E112 ensemble and, in order to obtain a robust estimate of the systematic errors associated with finite-volume effects, we also plan to use the other ensembles listed in \cref{tab:iso_EDI_FLAG}.

As already done in Ref.~\cite{DeSantis:2025qbb}, we adopt the mixed-action lattice setup introduced in~\cite{Frezzotti:2004wz} and described in full detail in the appendices of Ref.~\cite{ExtendedTwistedMassCollaborationETMC:2024xdf}. In this setup the action of the valence quarks is discretized in the so-called Osterwalder--Seiler (OS) regularization,
\begin{flalign}
&
S_\mathrm{OS}
=
\nonumber \\[8pt]
&
a^4\sum_x \bar{\psi}_q\left\{
\gamma_\mu \bar{\nabla}_\mu[U]-ir_q\gamma_5 \left(
W^\mathrm{cl}[U]+m_\mathrm{cr}
\right) +m_q
\right\}\psi_q \;,
\label{eq:OS_valence_quark_action}
\end{flalign}
where $q$ is the flavour index, the sum runs over the lattice points, $m_q$ is the bare quark mass, $m_\mathrm{cr}$ is the critical-mass counter-term and we refer to Refs.~\cite{Alexandrou:2018egz,ExtendedTwistedMass:2020tvp, ExtendedTwistedMass:2021qui,Finkenrath:2022eon} for the explicit definition of the covariant derivatives $\bar{\nabla}_\mu[U]$ and of the Wilson-Clover term $W^\mathrm{cl}[U]$, both depending on the gauge links $U_\mu(x)$. Valence and sea quarks have been simulated with the same value of $m_\mathrm{cr}$, tuned to restore chiral symmetry, and the bare masses $m_q$ of the up-down, strange and charm valence quarks have been tuned so that the corresponding renormalized masses match those of the sea quarks.
For each flavour $q$ we have two valence OS quark fields with opposite values of the Wilson parameters, $r_q=\pm 1$, and exploited this flexibility to choose, among the possible lattice discretizations of the required Euclidean correlators, the ones with the largest statistical signal-to-noise ratios (see Ref.~\cite{DeSantis:2025qbb} and below for more details).

In order to analyze our lattice data we use the bootstrap procedure explained in appendix~A of Ref.~\cite{DeSantis:2025qbb} which allows to combine results obtained from different simulations, to properly take into account statistical correlations and also to easily incorporate and appropriately treat (possibly correlated) systematic errors.

\subsection{
\label{sec:masstuning}
Tuning the $(m_i,m_f)$ heavy quark masses at fixed $r_{IF}$
}
On each gauge ensemble we calculate the decay rate for six pairs $(m_i,m_f)$ of initial and final heavy quark masses tuned to satisfy the condition in~\cref{eq:rIFcondition}.

In order to perform the tuning we study the non-perturbative dependence of the mass $M_{H}$ of heavy-strange pseudoscalar mesons over a wide range of heavy quark masses $m_h$. To interpolate the meson we use the following operator
\begin{flalign}
P_H(t,\vec{x}) = \sum_{\vec{y}}\bar \psi_s(t,\vec{x}) G^{N_\mathrm{sm}}_{t}(\vec{x},\vec{y}) \gamma_5 \psi_h(t,\vec{y}) \;,
\end{flalign}
where $\psi_s$ and $\psi_h$ are the strange and heavy quark fields, that we simulate with $r_h=-r_s$. In the previous expression $G_{t}
(\vec{x},\vec{y})$ is the Gaussian smearing operator
\begin{flalign}
G_{t}(\vec{x},\vec{y}) = \frac{1}{1+ 6\kappa}\left( \delta_{\vec{x},\vec{y}} + \kappa H_{t}(\vec{x},\vec{y})    \right)~,
\end{flalign}
with
\begin{flalign}
H_{t}(\vec{x}, \vec{y}) = \sum_{\mu=1}^{3}\left( \mathcal{U}_{\mu}(t,\vec{x})\delta_{\vec{x}+\hat{\mu},\vec{y}} + \mathcal{U}^{\dagger}_{\mu}(t,\vec{x}-\hat{\mu})\delta_{\vec{x}-\hat{\mu},\vec{y}}    \right)~,
\end{flalign}
and we have indicated with $\mathcal{U}_{\mu}(x)$ the APE-smeared links, defined as in Ref.~\cite{APE:1987ehd}. For this calculation, we employed the values $\kappa=0.5$ and fixed the number of smearing steps $N_\mathrm{sm}$ to obtain a smearing radius $a\sqrt{\frac{{N_\mathrm{sm}\kappa}}{{1+6\kappa}}}=0.18$~fm.

The meson mass $M_{H}(m_h)$ is extracted from the correlator
\begin{flalign}\label{eq:twopoint}
C(t)=\sum_{\vec x} T\bra{0} P_H(t,\vec x)\, P_H^\dagger(0) \ket{0}
\end{flalign}
which asymptotically ($0\ll t \ll T$) behaves as
\begin{flalign}
C(t)=\frac{R_{H}}{2 M_{H}}\, e^{-M_{H}t} + \cdots\;,
\label{eq:Csingleexp}
\end{flalign}
where the dots represent exponentially suppressed contributions.

As far as the initial heavy quark mass is concerned, we choose six values $m_i^{(n)}$, with $n=0,\cdots,5$, such that $m_i^{(n+1)}/m_i^{(n)}=1.28$. In correspondence of these values we compute the meson masses $M_I^{(n)}\equiv M_H(m_i^{(n)})$ and tune $m_i^{(0)}$ to satisfy the condition $M_I^{(0)}=M_{D_s}$.

\begin{figure}
    \centering
    \includegraphics[width=\linewidth]{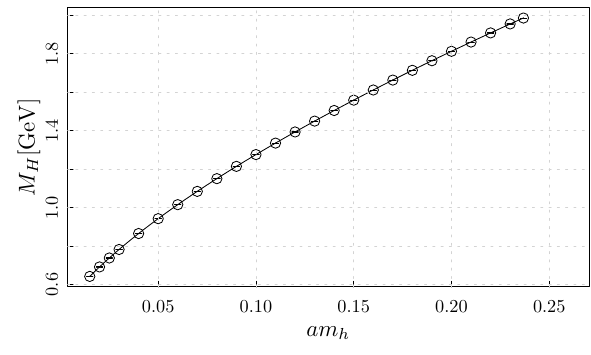}
    \caption{Pad\'e-fit of $M_H(m_h)$ on the B64 ensemble, see \cref{eq:masspade}.}
    \label{fig:padeb64}
\end{figure}
To determine the corresponding six $m_f^{(n)}$ final quark masses, we simulate $O(20)$ heavy quark masses $m_h$ and perform a $[2,2]$ Pad\'e-fit of the corresponding meson masses with the function
\begin{flalign}
    M_H(m_h)=\frac{n_0+n_1 m_h + n_2 m_h^2}{1+d_1 m_h + d_2 m_h^2}\;.
    \label{eq:masspade}
\end{flalign}
We found that this fit provides an excellent description of the numerical data on all ensembles (the measured data always deviate by less than $0.85 \sigma$ from the fit and the relative deviation between measurement and fit is always less than $2\cdot 10^{-4}$). The fit in the case of the B64 ensemble is shown in~\cref{fig:padeb64}. By using these fits, in correspondence of each $m_i^{(n)}$ we then solve numerically the equation $M_H(m_f^{(n)})=r_{IF}^\mathrm{phys} M_I^{(n)}$ and obtain the tuned sequence $(m_i,m_f)^{(n)}$ at fixed $M_F^{(n)}/M_I^{(n)}=r_{IF}^\mathrm{phys}=M_{D_s}/M_{\bar{B}_s}$.
The results are tabulated in \cref{tab:mesonmasses}.

\begin{table}[t!]
    \centering
    \begin{tabular}{lll}
    $n$ & $M_I [\si{\giga\electronvolt}]$ & $M_F [\si{\giga\electronvolt}]$\\
\hline \\[-2pt]

0 & $1.9674(12)$ & $0.72216(58)$\\
1 & $2.2790(22)$ & $0.83545(63)$\\
2 & $2.6570(34)$ & $0.97457(65)$\\
3 & $3.1170(49)$ & $1.14308(68)$\\
4 & $3.6713(74)$ & $1.34744(74)$\\
5 & $4.333(12)$ & $1.5954(12)$\\
\hline \\[-2pt]

    \end{tabular}
    \caption{The masses of the initial and final mesons.}
    \label{tab:mesonmasses}
\end{table}

\subsection{
\label{sec:fourpoint}
Calculation of the hadronic tensor correlators
}

\begin{table}[t!]
    \begin{tabular}{lcc}
    ensemble $\qquad$ & $Z_V$ & $Z_A$   \\[2pt]
    \hline
    \\[-2pt]
    B64         & $  0.706354(54) $ & $  0.74296(19) $\\
    C80         & $  0.725440(33) $ & $  0.75814(13) $\\
    D96         & $  0.744132(31) $ & $  0.77367(10) $\\
    [4pt]
    \hline

\end{tabular}
\caption{
The values of the renormalization constants $Z_A$ and $Z_V$ used in this work.
\label{tab:configsrenormalizationconstants}}
\end{table}

The four-point correlators from which we extract the amputated correlators $\hat H_{\mu\nu}(t;p,\vec \omega)$ and $\hat Z^{(p)}(t;\vec \omega)$ (see~\cref{sec:differentialrate})  are given by
\begin{flalign}
&
C_{\mu\nu}(t_\mathrm{snk},t,t_\mathrm{src},\vec \omega^2)
=
a^9\sum_{\vec x_\mathrm{snk},\vec x_\mathrm{src},\vec x}
e^{iM_{I} \vec \omega\cdot \vec x}\ \times
\nonumber \\[8pt]
&
\phantom{C_{\mu\nu}(t_\mathrm{snk},t,t_\mathrm{src})=}
T\bra{0} P_I(x_\mathrm{snk}) J_\mu^\dagger(x) J_\nu(0) P_I^\dagger(x_\mathrm{src})\ket{0}\;,
\end{flalign}
where $x=(t,\vec x)$, $x_\mathrm{snk}=(t_\mathrm{snk},\vec x_\mathrm{snk})$ and  $x_\mathrm{src}=(t_\mathrm{src},\vec x_\mathrm{src})$. For $J_\mu(x)$ we employ the so-called OS lattice discretization of the weak current, i.e.\
\begin{flalign}
J^{\mathrm{OS}}_\mu(x)=\bar \psi_f(x) \gamma_\mu(Z_V-Z_A\gamma_5) \psi_i(x)\,,
\qquad r_f=r_i\,.
\end{flalign}
The values of $Z_V$ and $Z_A$ used in this calculation are given in \cref{tab:configsrenormalizationconstants} and have been
determined in~\cite{ExtendedTwistedMass:2024nyi} following the Ward-identity method explained in Appendix B of Ref.~\cite{ExtendedTwistedMass:2022jpw}.

We compute $C_{\mu\nu}(t_\mathrm{snk},t,t_\mathrm{src})$ with spatial momenta $\vec \omega=(0,0,2\pi\theta/LM_{I})$ along the third spatial direction by using flavour-twisted boundary conditions~\cite{deDivitiis:2004kq} for the quark field $\psi_f$, i.e.\ $\psi_f(x+\hat z L) = \exp(2\pi \theta i)\psi_f(x)$.
For each $M_I$, we perform simulations with ten different values of $|\vec \omega|$,
\begin{flalign}
\frac{|\vec \omega|}{|\vec \omega|^\text{max}}
=\{0.1, 0.2, 0.3, 0.4, 0.5, 0.6, 0.7, 0.8, 0.9, 0.95\}\;.
\end{flalign}

The asymptotic behavior of the four-point correlator $C^{\mu\nu}(t_\mathrm{snk},t,t_\mathrm{src},\vec \omega^2)$ in the limits $T/2\gg t_\mathrm{snk} \gg t >0 \gg t_\mathrm{src} \gg -T/2$ is given by
\begin{flalign}
&
C^{\mu\nu}(t_\mathrm{snk},t,t_\mathrm{src},\vec \omega^2)
\nonumber \\[8pt]
&
=
\frac{R_I}{4\pi} e^{-M_I(t_\mathrm{snk}-t-t_\mathrm{src})}\,
\hat H_{\mu\nu}(t;p,\vec \omega)
+\cdots\;,
\label{eq:Cmunuasympt}
\end{flalign}
where $\hat H_{\mu\nu}(t;p,\vec \omega)$ is the amputated correlator defined in~\cref{eq:hmunut} and the dots represent again exponentially suppressed terms\footnote{We take the chance to correct a typo in Ref.~\cite{DeSantis:2025qbb}: the factor $m_{D_s}$ at denominator of Eqs.~(74) has to be removed and, consequently, the factor $m_{D_s}^2$ at numerator of Eq.~(75) has to be replaced with $m_{D_s}$.}. From the previous relation, by using the values of $R_I$ and $M_I$ extracted from $C(t)$ (see \cref{eq:Csingleexp}), we extract $\hat H_{\mu\nu}(t;p,\vec \omega)$ and then, by using~\cref{eq:Zgamma}, the three linear combinations $\hat Z^{(p)}(t;\vec \omega)$.
\begin{figure}
    \centering
    \includegraphics[width=\linewidth]{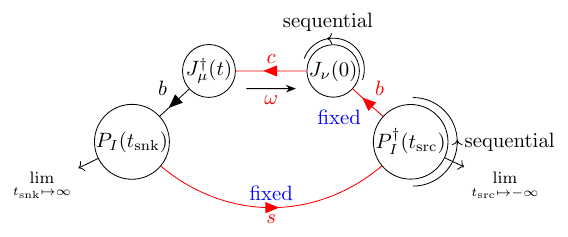}

    \includegraphics[width=\linewidth]{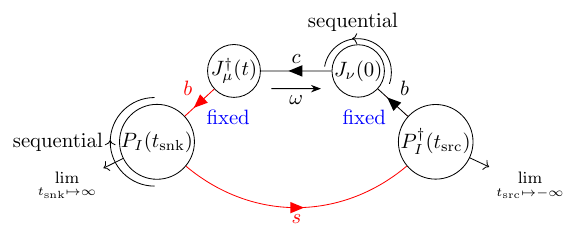}

    \caption{Sketch
    of the numerical implementation of the two strategies for the lattice calculation of $C_{\mu\nu}(t_\mathrm{snk},t,t_\mathrm{src})$. The upper diagram corresponds to the ``old strategy'', the one employed in all previous works on the subject. The lower diagram corresponds to the new strategy proposed and used in this work.
    The red lines correspond to one sequential quark propagator, the black lines to the other.
    In the end, both propagators are contracted.
    The old strategy requires $3 + N_J\cdot N_\omega=83$ inversions, while the new strategy requires $2 + N_J\cdot N_\omega + N_t$ inversions, with $N_J=4+4$ the number of currents that are inserted (vector plus axial components), $N_\omega=10$ the number of momenta that are simulated and $N_t$ the number of time separations that are considered.
    For the B64 ensemble, we have $N_t=32$, requiring $114$ inversions. On our largest ensemble, the E112, we will use $N_t=53$, requiring $135$ inversions.}
    \label{fig:BJmuvsBJnu}
\end{figure}
\begin{figure}
    \centering
    \includegraphics[width=\linewidth]{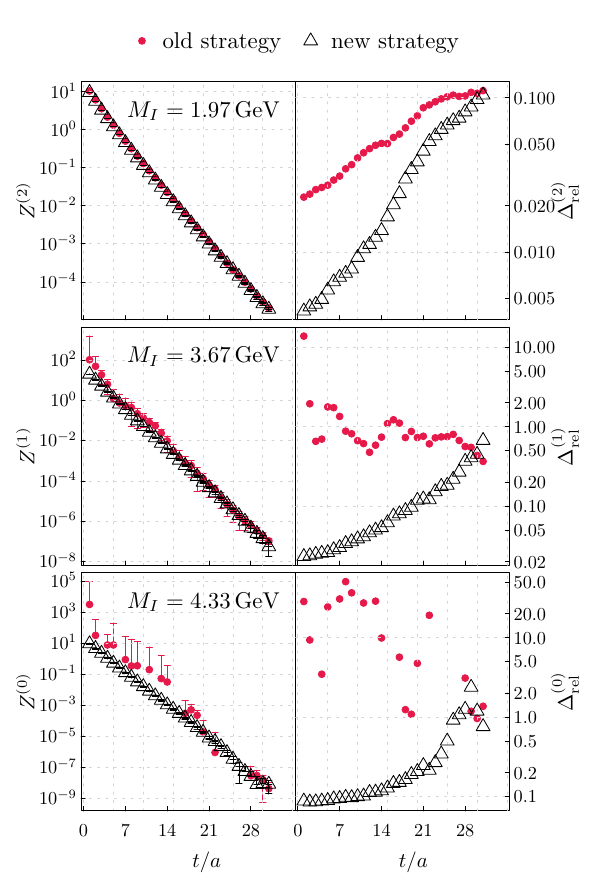}
    \caption{Comparison of the old and new simulation strategies.
    The left panels show the signals, while the right panels show the relative error, i.e.\ the noise-to-signal ratio.
    The red circles show the measurements done with the old strategy, while the black triangles those with the new strategy.
    All measurements were performed on the B64 ensemble with $|\vec \omega|=0.4 |\vec\omega|^\text{max}$.
    Points that do not have lower error bars are compatible with zero.
    The top panels correspond to the lightest heavy mass, $M_I=\qty{1.97}{\GeV}$, and show the $Z^{(2)}$ correlator. Here both simulation strategies lead to a good signal, but the relative error of the new strategy is smaller.
    The middle panel shows the correlator $Z^{(1)}$ for the second-heaviest mass, $M_I=\qty{3.67}{\GeV}$.
    Here, the signal is compatible with zero for the smallest time slices with the old simulation strategy, but resolved very well with the new strategy.
    Generally, the new strategy has a smaller relative error, and at large times the errors obtained with the two strategies are comparable.
    The bottom panels show the correlator $Z^{(0)}$ for the heaviest mass, $M_I=\qty{4.33}{\GeV}$.
    The signal is almost completely lost with the old simulation strategy.
    With the new strategy, the relative errors are larger than for smaller masses, but the signal is still resolved well. We observed the same pattern for all $Z^{0,1,2}$, at the spatial momenta and on all of the ensembles.}
    \label{fig:signaltonoisesimstrategies}
\end{figure}

An important technical result of this work, which represents a considerable improvement upon the current state-of-the-art, is a new numerical strategy for the lattice computation of $C^{\mu\nu}(t_\mathrm{snk},t,t_\mathrm{src},\vec \omega^2)$.

The new strategy, that we are now going to explain in details, is computationally more expensive than the one used in previous works~\cite{Gambino:2020crt,Gambino:2022dvu,Barone:2023tbl,DeSantis:2025yfm,DeSantis:2025qbb,Kellermann:2026sgp}, but allows to obtain significantly more precise results. This is particularly important in the heavy mass regime ($M_I>M_{D_s}$) where, by using the old strategy, we observed a rather unsatisfactory behaviour of the signal-to-noise ratio of $C^{\mu\nu}(t_\mathrm{snk},t,t_\mathrm{src},\vec \omega^2)$ by varying $t$.

We start by briefly recalling the old strategy, that we used and described
in Refs.~\cite{DeSantis:2025yfm,DeSantis:2025qbb}: $C^{\mu\nu}(t_\mathrm{snk},t,t_\mathrm{src},\vec \omega^2)$ is computed by keeping the time separations $0-t_\mathrm{src}$ and $t_\mathrm{snk}-t_\mathrm{src}$ fixed. In this strategy the operators $J_\nu(0)$ and $ P_I^\dagger(x_\mathrm{src})$ stay at fixed time separation while the operators $P_I(x_\mathrm{snk})$ and $ J_\mu^\dagger(x)$ move away from (get closer to) each other when $t$ decreases (increases).

As any Euclidean correlator, $C^{\mu\nu}(t_\mathrm{snk},t,t_\mathrm{src},\vec \omega^2)$ gets noisier when any of the three time separations between the four operators is increased. Actually, the signal-to-noise ratio rapidly deteriorates by increasing the time separations between $J_\nu(0)$ and $ P_I^\dagger(x_\mathrm{src})$, and/or between $P_I(x_\mathrm{snk})$ and $ J_\mu^\dagger(x)$. For this reason, the old strategy is not efficient for small values of $t$, that correspond to large $t_\mathrm{snk}-t$ separations.

To overcome this problem, in the new strategy we keep
the two time separations $t_\mathrm{snk}-t$ and $0-t_\mathrm{src}$ fixed and equal, for any value of $t$.

The numerical implementation of the two strategies is described in~\cref{fig:BJmuvsBJnu} while an example of the different behaviours of the signal-to-noise ratio as a function of $t$ is provided in~\cref{fig:signaltonoisesimstrategies}.

\begin{figure}
    \centering
    \includegraphics[width=\linewidth]{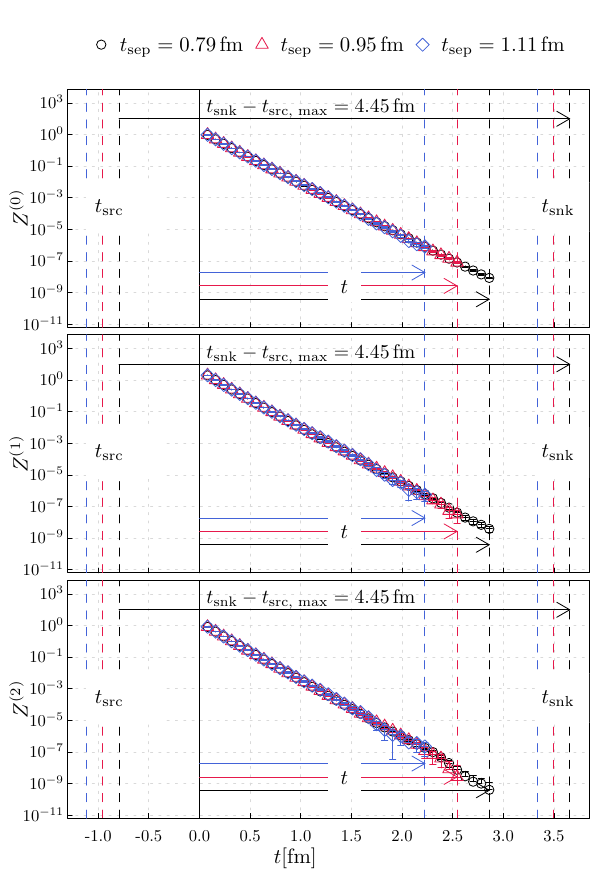}
    \caption{
    The signals of the $Z$-correlators for different $t_\text{sep}=0-t_\text{src}=t_\text{snk}-t$.
    The top, middle and bottom panels show the signal for $Z^{(0)}, Z^{(1)}$ and $Z^{(2)}$ respectively.
    The signals were measured on the B64 ensemble with $M_I=\qty{3.66}{\GeV}$ and $|\vec \omega|=0.4|\vec \omega|^\text{max}$.
    The black circles and black lines show the signal and setup for $t_\text{sep}=\qty{0.79}{\fm}$, the red triangle and lines correspond to $t_\text{sep}=\qty{0.95}{\fm}$ and the blue diamonds and lines correspond to $t_\text{sep}=\qty{1.11}{\fm}$.
    If no lower error bar is drawn, the signal is compatible with zero.
    The left lines show the  (fixed) location of $t_\text{src}$, the lines in the middle show the maximum value of $t$ that was measured and the right lines show the maximal value of $t_\text{snk}$ that was measured.}
    \label{fig:timeseparation}
\end{figure}

With the new strategy we have to ensure that we choose $t_\text{sep}=0-t_\text{src}=t_\text{snk}-t$ such that the asymptotic limits $t_\text{src}\to -\infty$, $t_\text{snk}\to \infty$, $T\to \infty$ are fulfilled.
To this end, we have computed the amputated correlators $\hat Z^{(p)}(t;\vec \omega)$ for several masses and momenta on the B64 ensembles for three different values of $t_\text{sep}$. In \cref{fig:timeseparation}
we show the results corresponding to $t_\text{sep}=10a_B=\qty{0.79}{\fm}$ in black, to $t_\text{sep}=12a_B=\qty{0.95}{\fm}$ in red and to $t_\text{sep}=14a_B=\qty{1.11}{\fm}$ in blue, where $a_B$ is the lattice spacing of the B64 ensemble (see \cref{tab:iso_EDI_FLAG}).
All signals agree within the statistical errors, and we choose the setup $t_\text{sep}=\qty{0.95}{\fm}$ for our further simulations, because it matches the region where the asymptotic behaviour of $C(t)$ sets in, see~\cref{eq:Csingleexp}, and it has sufficiently many points for the accuracy of the HLT spectral reconstruction.

\section{
\label{sec:resultswithhlt}
HLT Stability Analysis
}
In order to compute the decay rate we rely on the representation derived in~\cref{sec:differentialrate}, see in particular \cref{eq:lim_N_gamma,eq:gammalimit2}, and use the very same implementation of the HLT algorithm presented and discussed in full details in section VII of~\cite{DeSantis:2025qbb}. Here we do not repeat that presentation but we show examples of the stability analyses from which, on each ensemble, we extracted the $d \Gamma^{(p)}(\sigma)/d \vec \omega^2$ contributions to the differential decay rate.

We have considered two definitions of the smearing kernel $\Theta^{(p)}_\sigma(x)$ of \cref{eq:defthetap}, obtained by starting from the following two regularizations of the Heaviside step-function,
\begin{flalign}
\Theta_\sigma(x)=\frac{1}{1+e^{-\frac{x}{\sigma}}} \;,
\label{eq:sigmoid}
\end{flalign}
and
\begin{flalign}
\Theta_\sigma(x)=\frac{1+\mathrm{erf}\left(\frac{x}{s\sigma}\right)}{2}\; ,
\label{eq:erf}
\end{flalign}
where $s=2.5$ (see~\cite{DeSantis:2025qbb} for more details concerning this choice) and the error function is defined as
\begin{flalign}
\mathrm{erf}(x)=\frac{2}{\sqrt{\pi}}\int_0^x dt\, e^{-t^2}. \,
\end{flalign}
We call ``sigmoid kernel'' and ``error-function kernel'' the smooth functions $\Theta^{(p)}_\sigma(x)$ obtained multiplying by $x^{p}$ respectively \cref{eq:sigmoid,eq:erf}. We choose equal smearing parameters in physical units for all simulated values of $M_I$. In the case of the sigmoid kernel we considered the following ten values
\begin{flalign}
\sigma M_I=\qtylist[list-units = bracket, list-final-separator = {, }, list-open-bracket = \{, list-close-bracket = \}]{20; 40; 60; 80; 100; 120; 140; 160; 200}{\mega\electronvolt}\;,
\end{flalign}
while we did not use the two smallest values in the case of the error-function kernel.

\begin{figure}
    \centering
    \includegraphics[width=\linewidth]{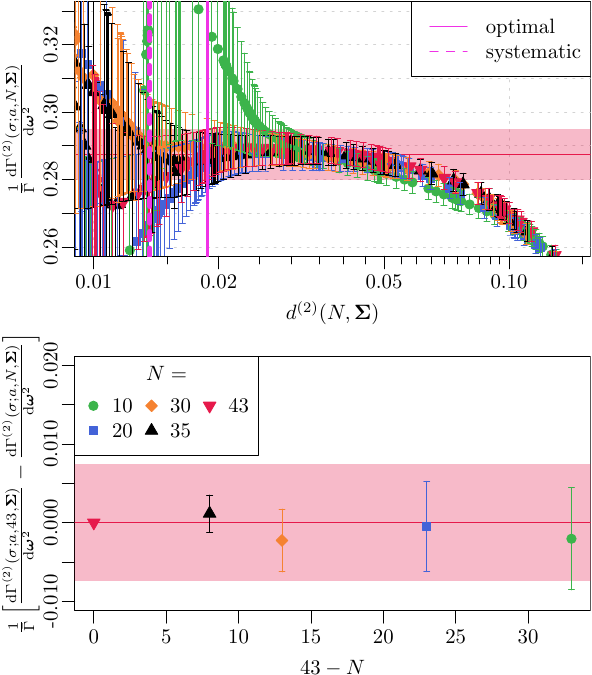}
    \caption{Stability analysis of the results for $d \Gamma^{(2)}(\sigma)/d \vec \omega^2$ obtained on the D96 ensemble, with $M_I=\qty{3.12}{\GeV}$, $|\vec \omega|=0.3|\vec\omega|^\text{max}$, $\sigma M_I=\qty{20}{\mega\electronvolt}$, $\alpha=0$ and the sigmoid smearing kernel.
    The top panel shows the result of the stability analysis, with the different colors and point shapes corresponding to different $N$. The shaded band represents the result $d \Gamma^{(p)}_\star(\sigma)/d \vec \omega^2$ extracted at $N=43$ in correspondence of the solid magenta line. The dashed magenta line represents the result  $d \Gamma^{(p)}_{\star\star}(\sigma)/d \vec \omega^2$ that was used to estimate the systematic error.
    The bottom panel shows the correlated differences between the results extracted in correspondence of the solid magenta line at $N=43$ and at the other values of $N$.}
    \label{fig:hltstabilitydifferentN}
\end{figure}

In \cref{fig:hltstabilitydifferentN}, which is analogous to Figure 8 of~\cite{DeSantis:2025qbb}, we show an example of the $N\to\infty$ limit of the results obtained with the HLT procedure. The stability analysis is the study of the dependence of the results for the differential decay rate  w.r.t.\ the accuracy of the reconstruction of the smearing kernel, see \cref{eq:stoneZ}, which is measured by the norm functionals
\begin{flalign}
&
A_\alpha^{(p)}[\vec g(N)] =
\int_{\omega^\mathrm{th}} d\omega\, e^{\omega_0 (a M_I) \alpha} \, \times
\nonumber \\[8pt]
&
\times\left[
\Theta_\sigma^{(p)}(\omega^\mathrm{max}-\omega_0)
-
\sum_{n=1}^{N} g_n(N)\, e^{-\omega_0 (a M_I) n}
\right]^2\,,
\label{eq:definitionnorm}
\end{flalign}
with $\alpha<2$.
In the HLT reconstruction, by varying the algorithmic parameters $\vec \Sigma=\{\omega^\mathrm{th},\alpha,\lambda\}$, where $\lambda$ is the so-called Backus--Gilbert trade-off parameter, one gets coefficients $\vec g^{(p)}(N;\vec \Sigma)$ which approximate the smearing kernels $\Theta_\sigma^{(p)}(\omega^\mathrm{max}-\omega_0)$ at different levels of accuracy. In the limit in which the kernel is exactly reconstructed, the quantity
\begin{flalign}
d^{(p)}\left(N;\vec \Sigma\right)=\sqrt{\frac{A_0^{(p)}\left[\vec g^{(p)}(N;\vec \Sigma)\right]}{A_0^{(p)}[\vec 0]}}\;,
\end{flalign}
vanishes exactly. The top panel of \cref{fig:hltstabilitydifferentN} shows the variation of the results for $d \Gamma^{(2)}(\sigma)/d \vec \omega^2$, computed on the D96 ensemble for $M_I=\qty{3.12}{\GeV}$, $|\vec \omega|=0.3|\vec\omega|^\text{max}$, $\sigma M_I=\qty{20}{\mega\electronvolt}$ and sigmoid smearing kernel, w.r.t.\ $d^{(2)}\left(N;\vec \Sigma\right)$ and for different values of $N$. The solid magenta line marks the value of $d^{(2)}\left(N;\vec \Sigma\right)$ at which we have extracted the central value, $d \Gamma^{(p)}_\star(\sigma)/d \vec \omega^2$, and the statistical error, $\Delta^{(p)}_\mathrm{stat}(\vec \omega,\sigma)$ (the shaded band), of our result. Indeed, as it can be seen, by further improving the accuracy of the kernel reconstruction the results are stable within their statistical errors that, actually, get rapidly very large in the $d^{(2)}\left(N;\vec \Sigma\right)\mapsto 0$ limit. Moreover, as shown in the bottom panel, in correspondence of the solid magenta line the correlated differences between the results obtained with $N=43$ and with smaller values of $N$ is compatible with zero within errors. From this analysis we conclude that the systematic error associated with the $N\to \infty$ limit is much smaller than the statistical error that we quote on our result. Nevertheless, to account for residual systematic errors, we consider a second point on the left of the solid magenta line, marked with the vertical dashed magenta line, from which we extract the result $d \Gamma^{(p)}_{\star\star}(\sigma)/d \vec \omega^2$, corresponding to a much more accurate reconstruction of the kernel\footnote{In our future work on the subject we will implement the improvement of the HLT algorithm which has been recently proposed in Ref.~\cite{Lupo:2026vdj}.
}.

To estimate the systematic errors we use the same data-driven procedure used in~\cite{DeSantis:2025qbb}.
In case we have two measurements $O_i$ and $O_j$ of an observable $O$, and we expect them to differ by an amount comparable to the systematic error, a general pull variable is defined as
\begin{flalign}
\mathcal{P}^{ij}_\mathrm{sys} = \frac{O_i-O_j}{\Delta_{ij}}\;,
\label{eq:pull}
\end{flalign}
where $\Delta_{ij}$ is a conservative estimate of the error of the difference $O_i-O_j$ (depending on the observable we consider either the error of one of the terms or the sum in quadrature of the errors of the two terms, see \cref{eq:PHLT}, \cref{eq:PCONT} and \cref{eq:PSIGMA}). We then estimate the systematic error by using the formula
\begin{flalign}
    &
    \Delta_\mathrm{sys}=
    \max_{ij}\left[
    \left|
    O_i-O_j
    \right|
    \mathrm{erf}\bigg(\frac{\mathcal{P}^{ij}_\mathrm{sys}}{\sqrt{2}}\bigg)
    \right]\; .
\label{eq:systematic_error}
\end{flalign}

\begin{figure}
    \centering
    \includegraphics[width=\linewidth]{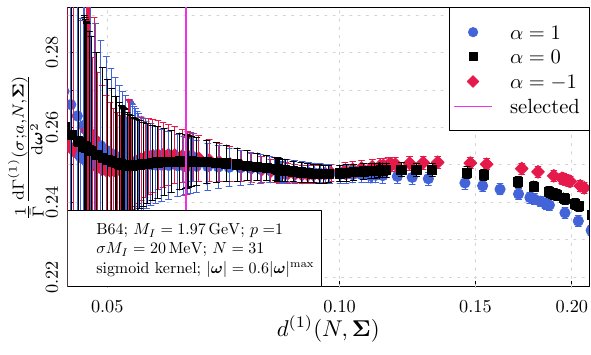}
    \includegraphics[width=\linewidth]{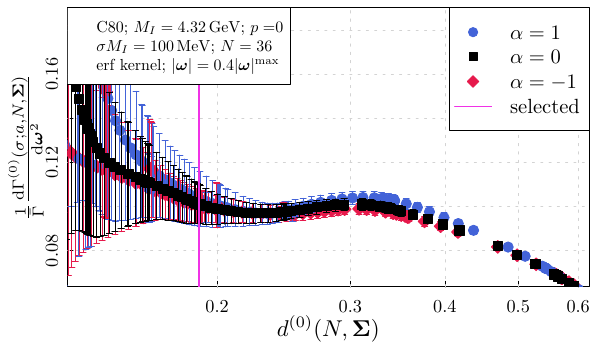}
    \includegraphics[width=\linewidth]{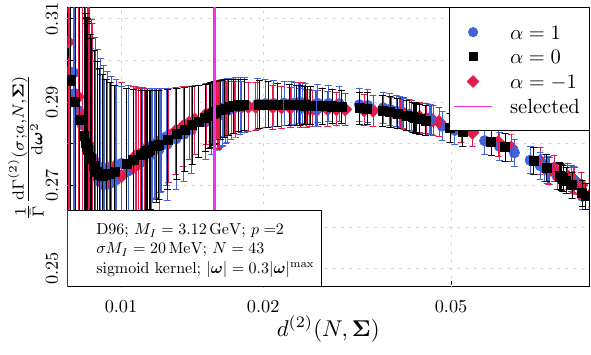}
    \caption{Further examples of stability plots.
    The blue circles correspond to the norm with $\alpha=1$, the black squares correspond to $\alpha=0$ and the red diamonds to $\alpha=-1$, see \cref{eq:definitionnorm}.
    The solid magenta line marks the value of $d^{(p)}\left(N;\vec \Sigma\right)$ where we extracted the $d \Gamma^{(p)}_{\star}(\sigma)/d \vec \omega^2$ results.
    Top panel: Computed on the B64 ensemble with $M_I=\qty{1.97}{\GeV}$, $|\vec \omega|=0.6|\vec\omega|^\text{max}$, $\sigma M_I=\qty{20}{\mega\electronvolt}$ and sigmoid smearing kernel.
    Middle panel: Computed on the C80 ensemble with $M_I=\qty{4.33}{\GeV}$, $|\vec \omega|=0.4|\vec\omega|^\text{max}$, $\sigma M_I=\qty{100}{\mega\electronvolt}$ and error-function smearing kernel.
    Bottom panel: Computed on the D96 ensemble with $M_I=\qty{3.12}{\GeV}$, $|\vec \omega|=0.3|\vec\omega|^\text{max}$, $\sigma M_I=\qty{20}{\mega\electronvolt}$ and sigmoid smearing kernel.}
    \label{fig:stabilitynorms}
\end{figure}

In the case of the HLT systematic error the pull variable is defined according to
\begin{flalign}\label{eq:PHLT}
\mathcal{P}^{(p)}_\mathrm{HLT}(\vec \omega,\sigma)=
\frac{1}{\Delta^{(p)}_\mathrm{stat}(\vec \omega,\sigma)\, \bar \Gamma}
\left(
    \frac{d\Gamma^{(p)}_\star\big(\sigma\big)}{d\vec \omega^2}
    -
    \frac{d\Gamma^{(p)}_{\star\star}\big(\sigma\big)}{d\vec \omega^2}
\right)
\,.
\end{flalign}
For the three ensembles, two kernels and ten momenta we have used for the analysis of the differential decay rate, we have analysed 9720 stability plots and determined the pull factors.
In 55\% of all cases, the pull factor has a modulus smaller than one, and in only 2.5\% of the cases the pull factor has a modulus larger than three.

Further examples of stability plots, showing the effect of using a different norm parameter $\alpha$, are provided in \cref{fig:stabilitynorms}.
For large values of $d^{(p)}\left(N;\vec \Sigma\right)$, corresponding to large deviations from the target kernel, there are differences larger than one standard deviation between the results corresponding to the different norms, especially visible in the top panel.
For smaller values of $d^{(p)}\left(N;\vec \Sigma\right)$, i.e.\ better kernel reconstruction, the different norms agree.
In all cases, we determine the central value of the decay rate at a point at which the different norms agree.
In all our analysis performed in this work we used the norm with $\alpha=0$ to determine the central values and the statistical errors.

\section{
\label{sec:results}
The $a\to 0$ and $\sigma \to 0$ limits and the numerical integral
}

After performing the $N\mapsto \infty$ limit with the HLT stability analysis, four more analysis steps are needed in order to get our final results for the total decay rate at fixed $M_I$, see \cref{eq:gammalimit2}.

In this paper, which is focused on the strategy of the calculation and where the result that we present is not the final one, we do not perform (and therefore do not discuss) the step of the infinite volume extrapolation. Our results have been obtained at $L\simeq 5.5$~fm and, in the following, we work under the assumption that finite volume effects are negligible. This is a reasonable assumption in the light of what we found in the $D_s\mapsto X\ell\nu$ case~\cite{DeSantis:2025yfm,DeSantis:2025qbb} since the $\bar B_s$ decay rate is expected to be a much more short-distance dominated quantity. Nevertheless, we will fill this gap in our forthcoming final publication on the subject.

The other three steps that have to be performed are the continuum limit $a \to 0$, the $\sigma \to 0$ limit and the numerical integration of the differential decay rate $d \Gamma(\sigma)/d \vec \omega^2$ over $\vec \omega^2$.
We call these steps C, S and I, respectively.
In principle, their orders are interchangeable, the only constraint is that the finite volume effects are accounted for before taking the $\sigma \to 0$ limit.
In the following subsections we investigate, separately, three different orders in which we performed these steps, namely CSI, SCI and SIC. By comparing the results obtained in these three analysis branches we checked that the systematic errors associated with these steps have been reliably quantified.

\subsection{The CSI order}
\label{sec:CSIresults}

In the CSI order, we calculate the total decay rate as
\begin{flalign}
&
\Gamma
=
\int_0^{(\vert\vec \omega\vert^\mathrm{max})^2} d\vec \omega^2\,
\sum_{p=0}^2 \lim_{\sigma\mapsto 0}\Bigg\{
\lim_{a \mapsto 0} \frac{\mathrm {d} \Gamma^{(p)} (\sigma, a)}{\mathrm{d}\vec\omega^2}
\Bigg\}
\label{eq:explanationCSI}
\end{flalign}
for each $M_I$.

For the continuum limit, we presently have three lattice spacings available. In a future paper we will analyze a finer lattice spacing, $a\simeq 0.05$~fm (see \cref{tab:iso_EDI_FLAG}), and perform a more systematic analysis of the continuum extrapolations, as we did in~\cite{DeSantis:2025qbb}. Here, by relying on the fact that in our lattice formulation physical quantities are automatically $O(a)$ improved, we simply perform a fit linear in $a^2$,
\begin{flalign}
&\frac{\mathrm {d} \Gamma^{(p)} (\sigma, a)}{\mathrm{d}\vec\omega^2}=\frac{\mathrm {d} \Gamma^{(p)} (\sigma)}{\mathrm{d}\vec\omega^2} + C \cdot a^2
\;.
\label{eq:contlimit}
\end{flalign}

\begin{figure}
    \centering
    \includegraphics[width=\linewidth]{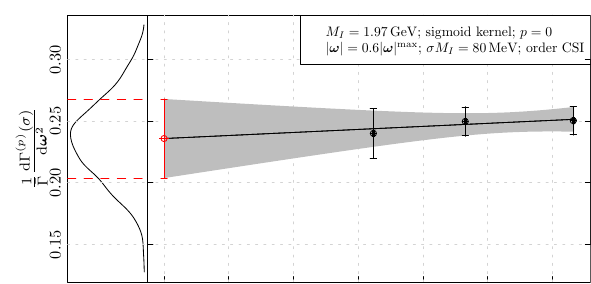}
    \includegraphics[width=\linewidth]{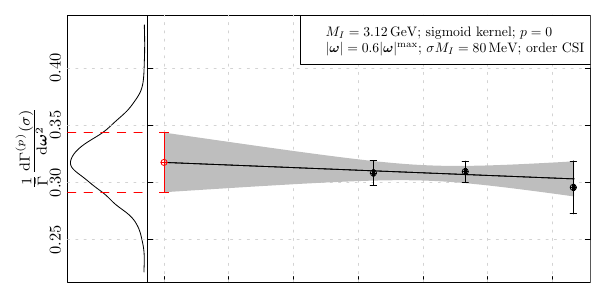}
    \includegraphics[width=\linewidth]{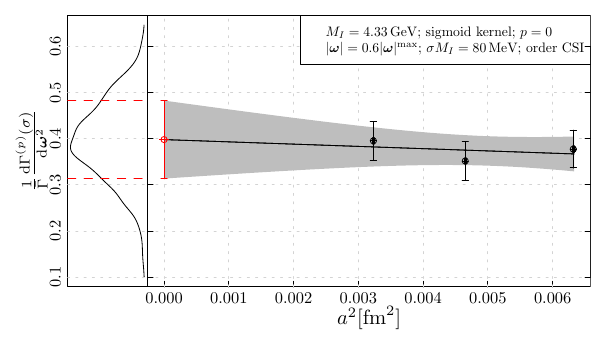}
    \caption{Continuum limit in the analysis order CSI with $\sigma M_I=\qty{80}{\mega\electronvolt}$, $|\vec \omega|=0.6|\vec \omega|^\text{max}$, $p=0$, using data obtained with the sigmoid smearing kernel.
    The top, middle, and bottom panels show $M_I=\qtylist[list-units = bracket, list-final-separator = {, }, list-open-bracket = \{, list-close-bracket = \}]{1.97;3.11;4.33}{\GeV}$, respectively.
    The left panel shows the density of the bootstrap samples of the continuum limit, with the red dashed lines denoting the statistical error.
    In the right panel, the black line and grey band correspond to the linear fit to the data, shown in black points.
    The continuum limit is shown as a red point, with two error bars corresponding to the statistical and total errors.
    Here the systematic error, which is added in quadrature to the statistical one, is so small that the error bars overlap.}
    \label{fig:CSIcontlimit}
\end{figure}

Examples of the continuum limit are shown in \cref{fig:CSIcontlimit}. Although, presently, we can only get a crude estimate of the systematic error of each single continuum limit (which is why we follow three different analysis orders), we consider the following pull factor
\begin{flalign}
&
\mathcal{P}^{(p)}_\text{cont}(\vec \omega)
\nonumber \\[8pt]
&=
\frac{1}{\Delta^{(p)}(\vec \omega, a=0)\, \bar \Gamma}
\left(
\frac{d\Gamma^{(p)}(a=0)}{d\vec \omega^2}
-
\frac{d\Gamma^{(p)}\big(a_\mathrm{min}^2\big)}{d\vec \omega^2}
\right)
\,.
\label{eq:PCONT}
\end{flalign}
and estimate the systematics as described in \cref{sec:resultswithhlt}, see \cref{eq:systematic_error}. In this analysis branch
we have performed over 3000 continuum limits, of which 77\% have a pull factor with modulus smaller than one and 3\% larger than three. The highest pull factors arise when analyzing the error-function kernel: for the sigmoid kernel, $89\%$ of all pull factors have a modulus smaller than one, and the highest pull factor is $2.5$.

After performing the continuum extrapolations we take the $\sigma \mapsto 0$ limits. To this end, we use the asymptotic formulae derived in section V of Ref.~\cite{DeSantis:2025qbb},
\begin{flalign}\label{eq:sigma_ansatz3}
\frac{d\Gamma^{(0),\mathrm{I}}(\sigma)}{d \vec \omega^2}
&=
C_0^{(0),\mathrm{I}}+C_1^{(0),\mathrm{I}}\sigma^2+C_2^{(0),\mathrm{I}}\sigma^4\;,
\nonumber \\[8pt]
\frac{d\Gamma^{(1),\mathrm{I}}(\sigma)}{d \vec \omega^2}
&=
C_0^{(1),\mathrm{I}}+C_1^{(1),\mathrm{I}}\sigma^2+C_2^{(1),\mathrm{I}}\sigma^4\;,
\nonumber \\[8pt]
\frac{d\Gamma^{(2),\mathrm{I}}(\sigma)}{d \vec \omega^2}
&=
C_0^{(2),\mathrm{I}}+C_1^{(2),\mathrm{I}}\sigma^4+C_2^{(2),\mathrm{I}}\sigma^6,
\nonumber \\[8pt]
\end{flalign}
where I=$\{$sigmoid, error-function$\}$ is the label associated with the two different smearing kernels. For each  contribution $\mathrm{d}\Gamma^{(p)}(\sigma)/\mathrm{d}\vec \omega^2$, we perform three different fits:
the first two correspond to separate polynomial extrapolations of the results obtained with the sigmoid and the error-function smearing kernels. The third fit is a combined extrapolation in which the coefficient of the constant term is the same for the two datasets, i.e.\  $C_0^{(p),\mathrm{sigmoid}}=C_0^{(p),\mathrm{error-function}}$. The three fits are then combined by using model averaging with the Akaike Information Criterion (AIC) to obtain our estimates  $\mathrm{d}\Gamma^{(p)}/\mathrm{d}\vec \omega^2$ of the physical differential decay rate. The formulae we use for the AIC are given in Eqs.~(97) to (99) of Ref.~\cite{DeSantis:2025qbb}.

\begin{figure}
    \centering
    \includegraphics[width=\linewidth]{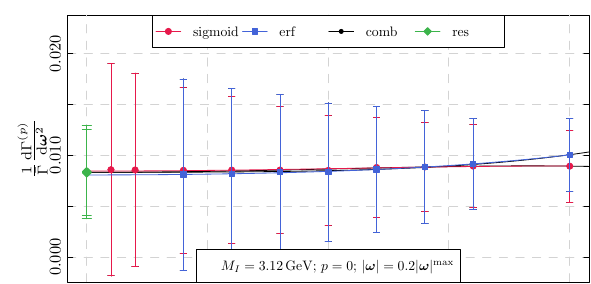}
    \includegraphics[width=\linewidth]{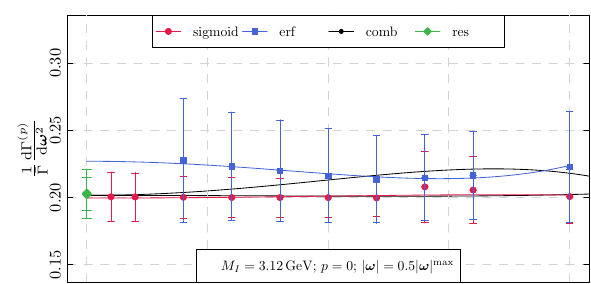}
    \includegraphics[width=\linewidth]{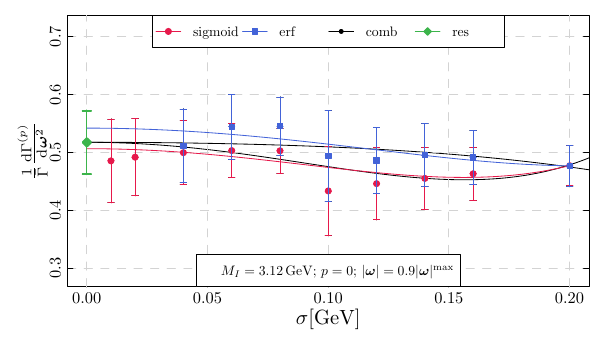}
    \caption{$\sigma \mapsto 0$ extrapolation of the differential decay rate for $M_I=\qty{3.12}{\GeV}$, $p=0$ in the CSI analysis order.
    The top, middle, and bottom panels show $|\vec \omega|=(\numlist[list-final-separator = {, }]{0.2; 0.5; 0.9})|\vec \omega|^\text{max}$, respectively.
    The red circles and line correspond to the data points and fit obtained with the sigmoid kernel, the blue squares and line correspond to the error-function kernel.
    The black lines correspond to the combined fit of the sigmoid and error-function data.
    The green diamond corresponds to the $\sigma \mapsto 0$ limit determined by combining the three fits according to their AIC weights.
    The two error bars represent the statistical and total error.
    }
    \label{fig:CSIsigma}
\end{figure}

Examples of $\sigma \mapsto 0$ extrapolations are shown in \cref{fig:CSIsigma}. We use the pull factor
\begin{flalign}\label{eq:PSIGMA}
\mathcal{P}^{(p)}_\sigma(\vec \omega)=
\frac{1}{\Delta^{(p)}(\vec \omega)\, \bar \Gamma}
\left(
\frac{d\Gamma^{(p)}}{d\vec \omega^2}
-
\frac{d\Gamma^{(p)}\big(\sigma^\mathrm{min}\big)}{d\vec \omega^2}
\right)
\,
\end{flalign}
to provide a quantitative measure of the overall quality of these fits. We have performed 720 $\sigma \mapsto 0$ extrapolations, with 96\% of the pull factors having a modulus smaller than one and no pull factors having a modulus larger than three.

To numerically perform the integral $\Gamma = \int_0^{(|\vec\omega|^\text{max})^2}\mathrm{d}\vec\omega^2 \frac{\mathrm{d} \Gamma}{\mathrm{d}\vec\omega^2}$, we interpolate the differential decay rate at the points that we measured.
We additionally consider the point $\frac{\mathrm{d} \Gamma(\vec{\omega}=0)}{\mathrm{d}\vec\omega^2}=0$ for our interpolation.
We consider three different interpolations/numerical-integration schemes:
an interpolation with cubic splines, which we then integrate exactly; a linear interpolation and the corresponding integration with the trapezoidal rule; an interpolation using piecewise parabolas and the corresponding Simpson's rule integration.

\begin{figure}
    \centering
    \includegraphics[width=\linewidth]{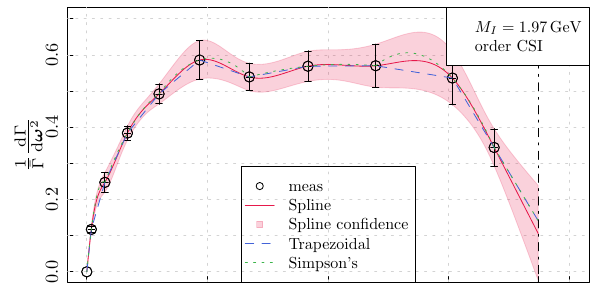}
    \includegraphics[width=\linewidth]{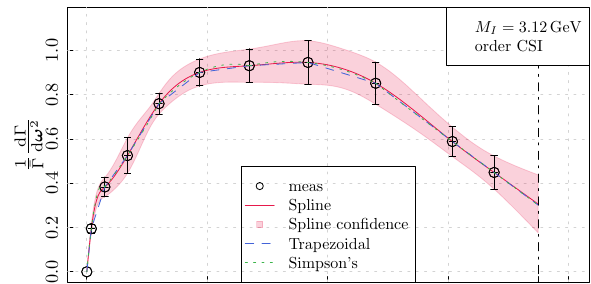}
    \includegraphics[width=\linewidth]{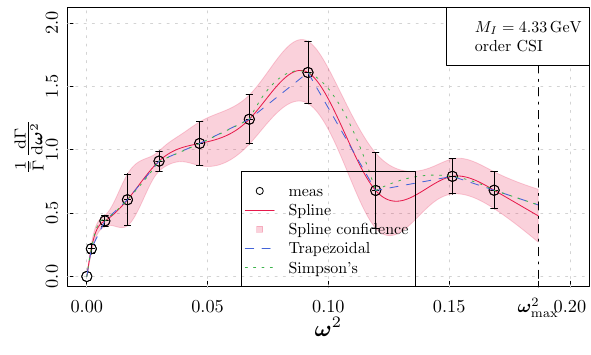}
    \caption{Graphical representation of the integral to determine $\Gamma$ in the CSI analysis order.
    The top, middle, and bottom panels show $M_I=\qtylist[list-units = bracket, list-final-separator = {, }, list-open-bracket = \{, list-close-bracket = \}]{1.97; 3.11; 4.33}{\GeV}$, respectively.
    The black points correspond to the lattice results after the continuum and $\sigma$ extrapolations, the red solid line is an interpolation of the points with cubic splines, the blue dashed line is the interpolation used for the trapezoid rule and the green dotted line is the interpolation used for Simpson's rule.
    The black dash-dotted line marks the end of the phase space.
    The shaded red band represents the confidence interval of the spline interpolation.}
    \label{fig:integralCSI}
\end{figure}

The integrals at $M_I=\qtylist[list-units = bracket, list-final-separator = {, }, list-open-bracket = \{, list-close-bracket = \}]{1.97; 3.11; 4.33}{\GeV}$ are shown in \cref{fig:integralCSI}.
For estimating the systematic error, we use \cref{eq:pull,eq:systematic_error} with $i, j \in $(Simpson, trapezoidal, spline) and for all values of $M_I$ the pull factor was below $0.4$.

\subsection{The SCI order}
\label{sec:SCIresults}

In the SCI order, we calculate the total decay rate as
\begin{flalign}
&
\Gamma
=
\int_0^{(\vert\vec \omega\vert^\mathrm{max})^2} d\vec \omega^2\,
\sum_{p=0}^2 \lim_{a\mapsto 0}\Bigg\{
\lim_{\sigma \mapsto 0} \frac{\mathrm {d} \Gamma^{(p)} (\sigma, a)}{\mathrm{d}\vec\omega^2}
\Bigg\}
\label{eq:explanationSCI}
\end{flalign}
for each $M_I$.

\begin{figure}
    \centering
    \includegraphics[width=\linewidth]{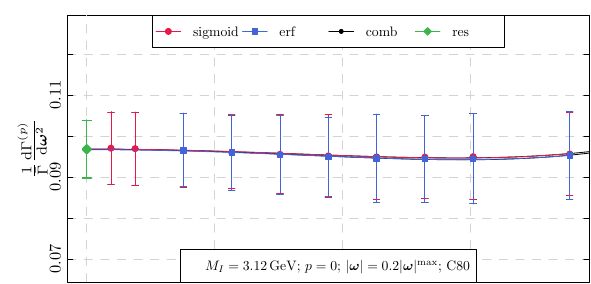}
    \includegraphics[width=\linewidth]{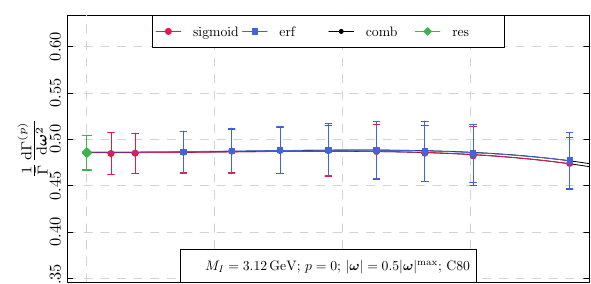}
    \includegraphics[width=\linewidth]{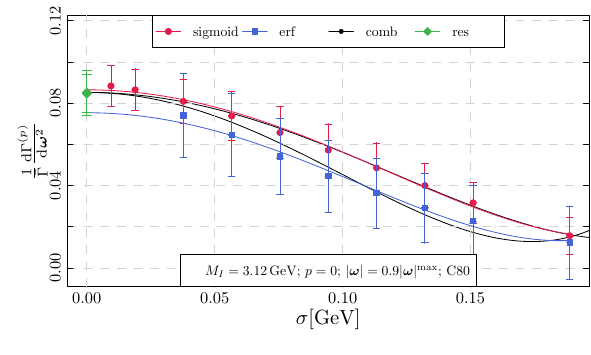}
    \caption{$\sigma \mapsto 0$ extrapolation of the differential decay rate for $M_I=\qty{3.12}{\GeV}$, $p=0$ for the C80 ensemble in the order SCI.
    The top, middle, and bottom panels show $|\vec \omega|=(\numlist[list-final-separator = {, }]{0.2; 0.5; 0.9})|\vec \omega|^\text{max}$, respectively.
    The red circles and line correspond to the data points and fit obtained with the sigmoid kernel, the blue squares and line correspond to the error-function kernel.
    The black lines correspond to the combined fit of the sigmoid and error-function data.
    The green diamond corresponds to the $\sigma \mapsto 0$ limit determined by combining the three fits according to their AIC weights.
    The two error bars represent the statistical and total error.
    }
    \label{fig:SCIsigma}
\end{figure}

The $\sigma \mapsto 0$ extrapolations have been performed with the asymptotic formulae given in \cref{eq:sigma_ansatz3}.
Example  extrapolations are shown in \cref{fig:SCIsigma}. For the SCI order we performed 2160 $\sigma \mapsto 0$ extrapolations, with 93\% of the pull factors (see \cref{sec:CSIresults}) having a modulus smaller than one and 1\% larger than three.

\begin{figure}
    \centering
    \includegraphics[width=\linewidth]{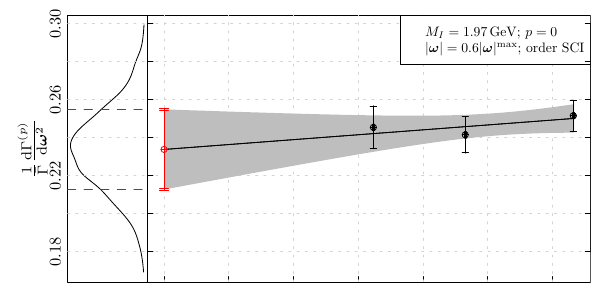}
    \includegraphics[width=\linewidth]{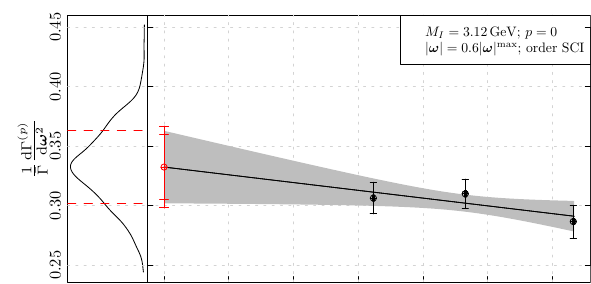}
    \includegraphics[width=\linewidth]{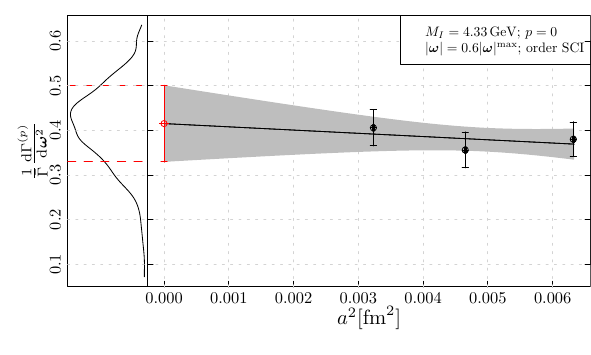}
    \caption{Continuum limit in the analysis order SCI with $|\vec \omega|=0.6|\vec \omega|^\text{max}$, $p=0$, using data obtained with the AIC criterion after the $\sigma \mapsto 0$ limit.
    The top, middle, and bottom panels show $M_I=\qtylist[list-units = bracket, list-final-separator = {, }, list-open-bracket = \{, list-close-bracket = \}]{1.97; 3.11; 4.33}{\GeV}$, respectively.
    The left panel shows the density of the bootstrap samples of the continuum limit, with the red dashed lines denoting the statistical error.
    In the right panel, the black line and grey band correspond to the linear fit to the data, shown in black points.
    The continuum limit is shown as a red point, with two error bars corresponding to the statistical and total errors.}
    \label{fig:SCIcont}
\end{figure}

The continuum limits have been performed again using a linear in $a^2$ ansatz and the associated systematic errors have been estimated by using  \cref{eq:PCONT}.
Example continuum extrapolations are shown in \cref{fig:SCIcont}.
We performed 180 continuum limits in the SCI order. Of these, 90\% had a pull factor with modulus less than one, and no extrapolation had a pull factor with modulus larger than three.

\begin{figure}
    \centering
        \includegraphics[width=\linewidth]{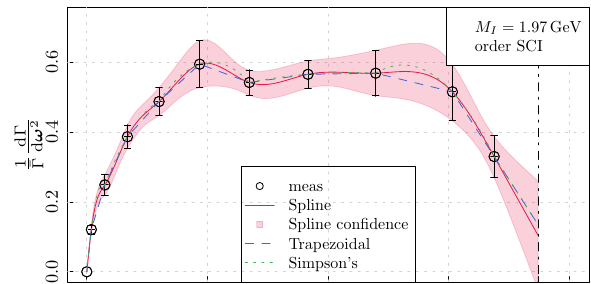}
        \includegraphics[width=\linewidth]{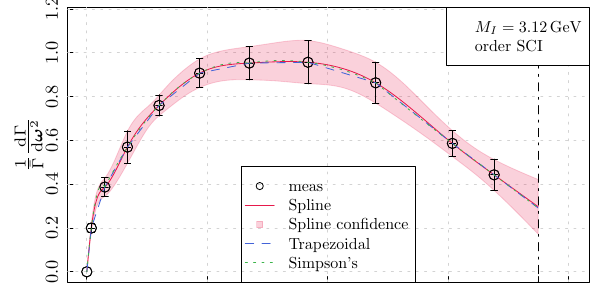}
        \includegraphics[width=\linewidth]{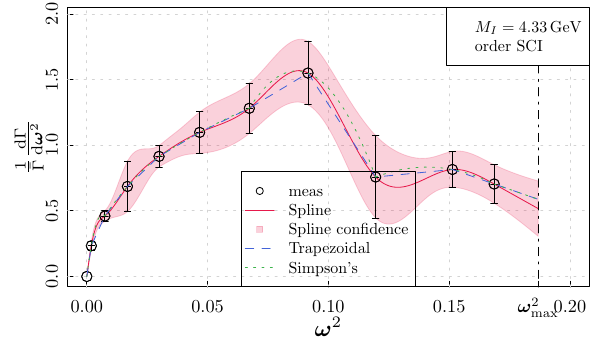}
    \caption{Graphical representation of the integral to determine $\Gamma$, for the SCI analysis order.
    The top, middle, and bottom panels show $M_I=\qtylist[list-units = bracket, list-final-separator = {, }, list-open-bracket = \{, list-close-bracket = \}]{1.97; 3.11; 4.33}{\GeV}$, respectively.
    The black points correspond to the lattice results after the continuum and $\sigma$ extrapolations, the red solid line is an interpolation of the points with cubic splines, the blue dashed line is the interpolation used for the trapezoid rule and the green dotted line is the interpolation used for Simpson's rule.
    The black dash-dotted line marks the end of the phase space.
    The shaded red band represents the confidence interval of the spline interpolation.}
    \label{fig:integralSCI}
\end{figure}

The integrals at $M_I=\qtylist[list-units = bracket, list-final-separator = {, }, list-open-bracket = \{, list-close-bracket = \}]{1.97; 3.11; 4.33}{\GeV}$ are shown in \cref{fig:integralSCI}.
They are performed as explained in \cref{sec:CSIresults}. We also considered the pull factors introduced in  \cref{sec:CSIresults} and all of them had a modulus
smaller
than $0.3$.

\subsection{The SIC order}
\label{sec:SICresults}

In the SIC order, we calculate the total decay rate as
\begin{flalign}
&
\Gamma
=
\lim_{a\mapsto 0}
\int_0^{(\vert\vec \omega\vert^\mathrm{max})^2} d\vec \omega^2\,
\sum_{p=0}^2 \Bigg\{
\lim_{\sigma \mapsto 0} \frac{\mathrm {d} \Gamma^{(p)} (\sigma, a)}{\mathrm{d}\vec\omega^2}
\Bigg\}
\label{eq:explanationSIC}
\end{flalign}
for each $M_I$.

The results of the limit $\sigma \mapsto 0$ are identical to those of the SCI order, see \cref{sec:SCIresults}.

\begin{figure}
    \centering
        \includegraphics[width=\linewidth]{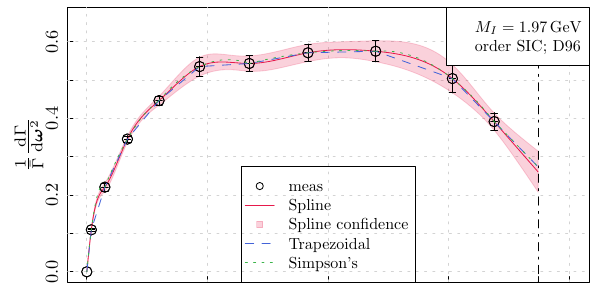}
        \includegraphics[width=\linewidth]{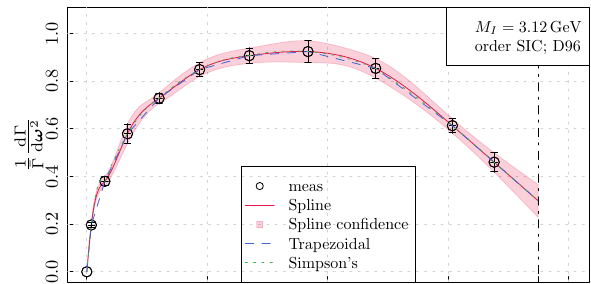}
        \includegraphics[width=\linewidth]{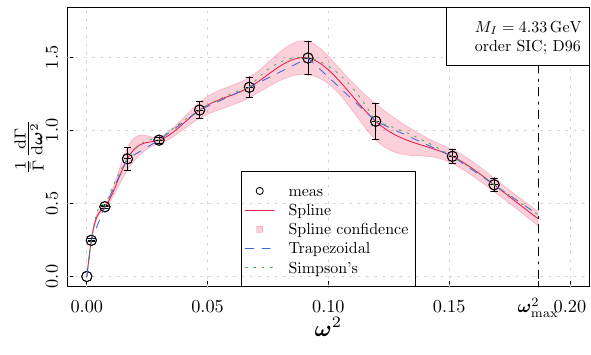}
    \caption{Graphical representation of the integral for determining $\Gamma$, in the SIC analysis order and on the D96 ensemble.
    The top panel shows the data for $M_I=\qty{1.97}{\giga\electronvolt}$, the middle panel for $M_I=\qty{3.11}{\giga\electronvolt}$ and the bottom panel for $M_I=\qty{4.33}{\giga\electronvolt}$.
    The black points correspond to the lattice data after the $\sigma$ extrapolations, the red solid line is an interpolation of the points with cubic splines, the blue dashed line is the interpolation used for the trapezoid rule and the green dotted line is the interpolation used for Simpson's rule.
    The black dash-dotted line marks the end of the phase space.
    The shaded red band represents the confidence interval of the spline interpolation.}
    \label{fig:integralSIC}
\end{figure}

The integrals at $M_I=\qtylist[list-units = bracket, list-final-separator = {, }, list-open-bracket = \{, list-close-bracket = \}]{1.97; 3.11; 4.33}{\GeV}$ for the D96 ensemble are shown in \cref{fig:integralSIC}.
The calculation of the integrals and the estimation of the systematic errors is performed analogously to \cref{sec:CSIresults}.
All 18 integrals we calculated for the analysis order SCI had a pull factor with modulus smaller than $0.6$.

\begin{figure}
    \centering
    \includegraphics[width=\linewidth]{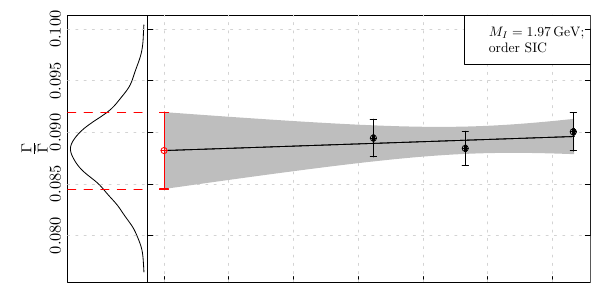}
    \includegraphics[width=\linewidth]{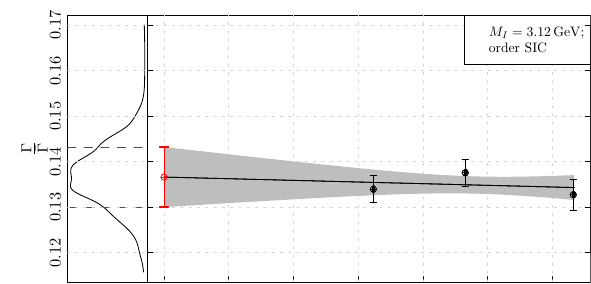}
    \includegraphics[width=\linewidth]{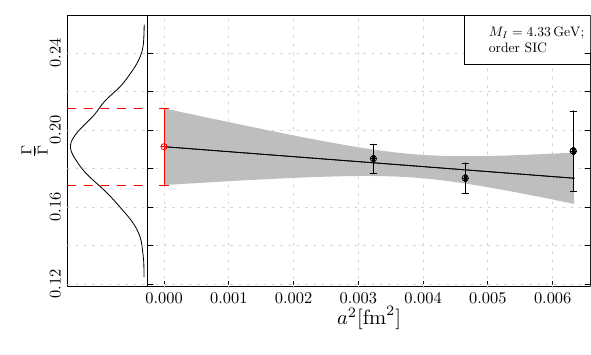}
    \caption{Continuum limit in the analysis order SIC.
    The top panel shows the data for $M_I=\qty{1.97}{\giga\electronvolt}$, the middle panel for $M_I=\qty{3.11}{\giga\electronvolt}$ and the bottom panel for $M_I=\qty{4.33}{\giga\electronvolt}$.
    The left panel shows the density of the bootstrap samples of the continuum limit, with the red dashed lines denoting the statistical error.
    In the right panel, the black line and grey band correspond to the linear fit to the data, shown in black points.
    The continuum limit is shown as a red point, with two error bars corresponding to the statistical and total errors.}
    \label{fig:SICcont}
\end{figure}

Examples of continuum extrapolations in the order SIC are given in \cref{fig:SICcont}, with the ansatz  again being linear in $a^2$, but this time applied to the total rate $\Gamma$ at fixed $M_I$.
The estimation of the systematic error is performed analogously to \cref{sec:CSIresults}.
In the analysis with order SIC, we performed 6 continuum limits. All pull factors had a modulus smaller than one.

\begin{figure}
    \centering
    \includegraphics[width=\linewidth]{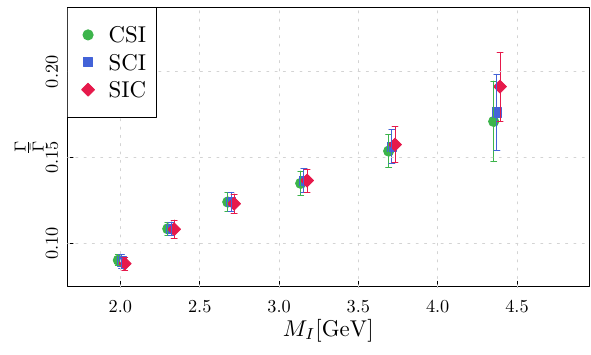}
    \caption{Total decay rate for all masses and all analysis orders.
    The green points correspond to analysis order CSI, the blue squares to order SCI and the red diamonds to order SIC.
    The markers of orders SCI and SIC were shifted along the x-axis to make the figure more legible.}
    \label{fig:compareanalysisorders}
\end{figure}

\subsection{Comparison of the results of the different orders}
\label{sec:ComparisonOrderResults}

We show the comparison of the results for the total decay rate as a function of $M_I$, obtained in the three different analysis orders, in \cref{fig:compareanalysisorders}.

The main message that we get from this branched analysis is that, within the quoted total errors, the results are fully compatible. This is reassuring evidence of the fact that, overall, our estimates of the systematic errors associated with the $a\mapsto 0$ and $\sigma\mapsto 0$ extrapolations and with the numerical integration of the differential decay rate are solid.

It should be noted, however, that in all analysis orders we adopted a rather simple strategy to perform the continuum extrapolations: we only have three lattice spacings and we opted for a linear fit in $a^2$ always including all points.
The control over the continuum limit is \emph{crucially important} in any $b$-physics lattice calculation, including the present one.
Nevertheless we consider it very reassuring that, despite the differences in the differential decay rates between the CSI, SCI and SIC orders (see \cref{fig:integralCSI,fig:integralSCI,fig:integralSIC}), the final results are well compatible within estimated uncertainties (see \cref{fig:compareanalysisorders}).
This is of particular relevance because the CSI and SCI orders require many more continuum extrapolations w.r.t.\ the single extrapolation required in the SIC order at fixed $M_I$.
Our current results make us very confident that, once we will add the ensemble at the finest value of the lattice spacing (the E112 one in \cref{tab:iso_EDI_FLAG}), we will be able to substantially reduce the total error by getting a more robust estimate of the systematic uncertainties associated with the continuum extrapolations, especially by adopting the SIC order in which the number of required $a\mapsto 0$ extrapolations is minimal. In any case, our current result can be considered sufficiently solid to be used in phenomenological analyses.

\section{The $M_I \mapsto M_{\bar B_s}$ interpolation}
\label{sec:interpolation}

In this section we describe the last step of the analysis: the interpolation of our lattice results, obtained in the continuum limit at six values of $M_I\le 4.3$~GeV, to the physical point $M_I=M_{\bar B_s}$. We perform the interpolation by using the remarkable fact, discussed in \cref{sec:OPE}, that the leading order of the OPE expansion in powers of $\LamQCD/M_I$ is purely partonic and, therefore, it does not require any non-perturbative input to be computed.

In order to incorporate the partonic result $\Gamma_\mathrm{part}$ of \cref{eq:partonicres} into our analysis procedure we follow two complementary strategies. In the first strategy, which we call the ``fitting-function'' approach, we choose a fitting function which reduces to $\Gamma_\mathrm{part}$ in the $M_I\mapsto \infty$ limit and that is sufficiently flexible to describe our non-perturbative lattice data also in the region around $M_I\sim 2$~GeV. In the second strategy, that we call the ``functional-space'' approach, we perform the interpolation by using Gaussian Processes. In this case, we obtain the posterior distribution of $\Gamma$ as a function of $M_I$ by starting from $\Gamma_\mathrm{part}$ as the input model and by choosing a prior distribution such that the constraints on the posterior at $M_I=M_{\bar B_s}$ come from the lattice data and from the fact that  $\Gamma=\Gamma_\mathrm{part}$ in the $M_I\mapsto \infty$ limit. The two approaches are described in detail, and then compared, in the following subsections.

\subsection{The fitting-function approach}
\label{sec:frequentist}

\begin{figure}
    \centering
    \includegraphics[width=\linewidth]{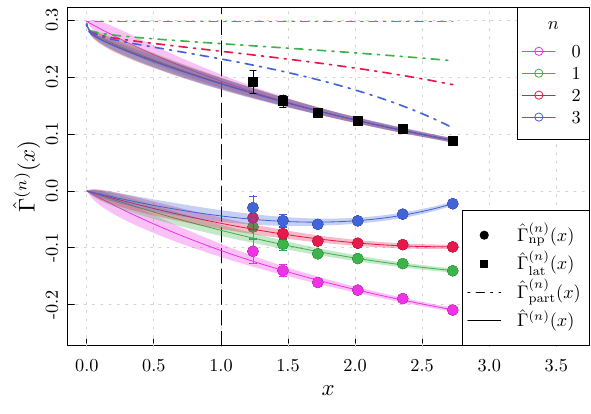}
    \caption{The interpolation of the decay rate at the physical point $x=1$ performed by using \cref{eq:freqfitfunc} with $\mu_s=\frac{M_{\bar{B}_s}}{2x}$ and for different values of the perturbative order $n$, for analysis order SIC.    The dashed lines at the top represent $\hat{\Gamma}_\mathrm{part}^{(n)}(x)$. The black squares in the middle represent our non-perturbative lattice results for the total decay rates, obtained with the analysis order SIC. The coloured circles at the bottom are obtained by subtracting $\hat{\Gamma}_\mathrm{part}^{(n)}(x)$ from the lattice data, with the lines and shaded bands representing the fit to the data according to $\hat{\Gamma}_\mathrm{np}^{(n)}(x)$ and the confidence interval of the fit. The shaded bands in the middle represent $\hat{\Gamma}^{(n)}(x)$, i.e. the sum of the shaded bands at the bottom and the dashed lines at the top. The dashed vertical line represents the physical point.}
    \label{fig:padeperturbativeorders}
\end{figure}

\begin{figure}
    \centering
    \includegraphics[width=\linewidth]{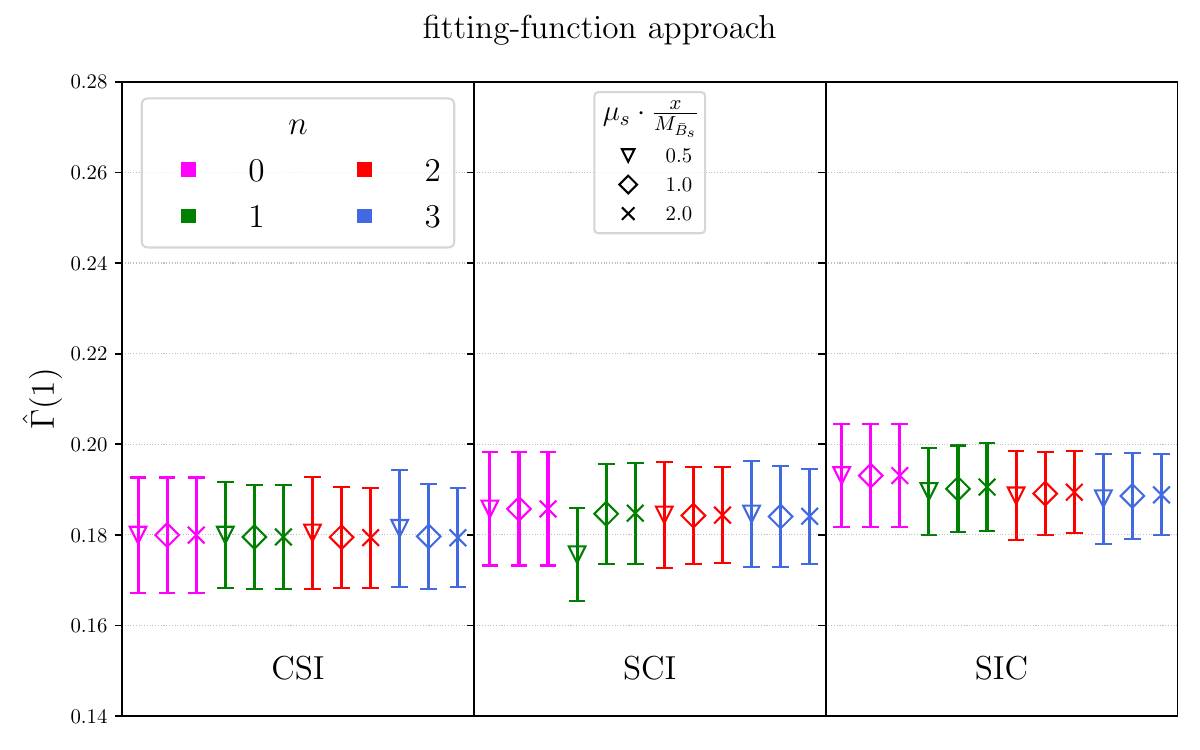}
    \caption{The result of the interpolation of $\hat{\Gamma}$ at the physical point by using the $\hat{\Gamma}^{(n)}$ parametrization of \cref{eq:freqfitfunc}.
    The magenta (leftmost) points in each panel correspond to $n=0$, the green (center-left) points to $n=1$, the red (center-right) points to $n=2$ and the blue (rightmost) points correspond to $n=3$.
    The triangle, diamond and cross markers correspond to $\mu_s(x)= \left\{\frac{1}{2},1,2\right\}\, \frac{M_{\bar B_s}}{x}$, respectively.
    The left, center and right panels correspond to the analysis orders CSI, SCI and SIC, respectively.
    The results at $n=0$ do not depend on the scale of $\mu_s$ by construction, as explained in \cref{sec:OPE}.}
    \label{fig:frequentistcomparison}
\end{figure}

We express the normalized total decay rate as a function of
\begin{flalign}
x=\frac{M_{\bar B_s}}{M_I}\;,
\qquad
\hat \Gamma(x) = \frac{\Gamma(x)}{\bar \Gamma(x)}\;,
\end{flalign}
so that the physical point is at $x=1$, and consider the following fitting functions
\begin{flalign}
\hat{\Gamma}^{(n)}(x)
= \hat{\Gamma}_\mathrm{part}^{(n)}(x)
+\hat{\Gamma}_\mathrm{np}^{(n)}(x) \;,
\label{eq:freqfitfunc}
\end{flalign}
where $\hat{\Gamma}_\mathrm{part}^{(n)}(x)$ is the partonic contribution and $\hat{\Gamma}_\mathrm{np}^{(n)}(x)$ the non-perturbative one. To obtain the partonic contribution we start from \cref{eq:partonicres}, which we evaluate at
\begin{flalign}
\mu_s(x)= \left\{\frac{1}{2},1,2\right\}\, \frac{M_{\bar B_s}}{x}\;,
\end{flalign}
and consider different orders of the perturbative expansion, namely $n=0,\cdots,3$, according to
\begin{flalign}
&
\hat{\Gamma}_\mathrm{part}^{(n)}(x)
=
\frac{\pi}{4} \sum_{k=0}^n \frac{\as^k(\mu_s(x))}{\pi^k} X_k\;.
\end{flalign}
We parametrize the non-perturbative contribution by the following rational function,
\begin{flalign}
\hat{\Gamma}_\mathrm{np}^{(n)}(x) = x\, \frac{a_0^{(n)}+a_1^{(n)}\, x}{1+b_0^{(n)}\, x}\;,
\label{eq:gammanp}
\end{flalign}
which, being proportional to $1/M_I$, is consistent with the OPE expansion and with the exact theoretical constraint
\begin{flalign}
\hat{\Gamma}^{(n)}(0)
=
\hat{\Gamma}_\mathrm{part}^{(n)}(0)
= \frac{\pi}{4}\, X_0\;,
\end{flalign}
that, in this way, is automatically enforced in our fits. Notice that the dependence of the coefficients in \cref{eq:gammanp} upon $n$ is inherited by the $\hat{\Gamma}_\mathrm{part}^{(n)}$ contribution in \cref{eq:freqfitfunc}. Indeed, provided that our parametrization of the non-perturbative contributions is sufficiently general, we must then find that the final result for $\hat{\Gamma}$ does not depend on $n$ within errors.

In \cref{fig:padeperturbativeorders} we show the results of our interpolations corresponding to the choice $\mu_s=M_I/2$, the analysis order SIC and to the different orders of the perturbative expansion: $n=0$ magenta, $n=1$ green, $n=2$ red, $n=3$ blue.

There is a \emph{very important} remark that is in order here. By using the parametrization of \cref{eq:freqfitfunc} we are using the perturbative results $\hat{\Gamma}_\mathrm{part}^{(n)}(x)$ \emph{way beyond} the range in which perturbation theory in the on-shell scheme is expected to be reliable (see the discussion in \cref{sec:OPE}). This is intentional, for the reason we are now going to explain, but in no way the results of our fits can be used to deduce information concerning the relative size of the perturbative and the non-perturbative contributions for $x\gg 0$. Indeed, as it is evident by looking at the different dashed lines in \cref{fig:padeperturbativeorders}, in the non-asymptotic regime ($x\ge 1$) the partonic results get very different by increasing the perturbative order and certainly do not converge. On the other hand, the chosen functional form for the non-perturbative contribution $\hat{\Gamma}_\mathrm{np}^{(n)}(x)$ is \emph{sufficiently general} to compensate for these differences. This is evident from the fact that the different bands, corresponding to the fits performed using $\hat{\Gamma}^{(n)}(x)$ at different values of $n$, are fully compatible in the non-perturbative region (see
\cref{fig:frequentistcomparison}) and provide an excellent description of our lattice data for any considered value of $n$. In fact, renormalization group equations imply that the decay rate $\hat{\Gamma}$ must have a logarithmic dependence w.r.t.\ $M_I$. The logarithmic behaviour in the perturbative region is taken into account by $\hat{\Gamma}_\mathrm{part}^{(n)}(x)$. Concerning the non-perturbative region, logarithms are not explicitly present in $\hat{\Gamma}_\mathrm{np}^{(n)}(x)$. On the other hand, the fact that the chosen rational parametrization of $\hat{\Gamma}_\mathrm{np}^{(n)}(x)$ is able to compensate the very different behaviours of $\hat{\Gamma}_\mathrm{part}^{(n)}(x)$ for different values of $n$ is a strong piece of evidence of the high flexibility of our fitting functions, that can actually provide a robust parametrization of any theoretically-reasonable behaviour of $\hat{\Gamma}$ as a function of $M_I$ in the non-perturbative region, and so at the physical point $x=1$.

\subsection{The functional-space approach}
\label{sec:bayesian}

In this section we address the problem of the interpolation of our lattice results for $\hat{\Gamma}(x)$ to the physical point, $x=1$, by using the Bayesian approach of Gaussian Processes, see e.g.\ Refs.~\cite{10.1093/gji/ggz520,Rasmussen,DelDebbio:2024lwm}.

The theory of Gaussian Processes is well consolidated and commonly used to address inference problems in many research fields. It is not so commonly used, though, to analyze lattice simulations' data and, for this reason, the following discussion is intentionally very detailed.

In the functional-space approach, which is dual and complementary to the fitting-function one, the problem is formulated from the  probabilistic perspective. The goal is that of obtaining the probability distribution over the functional space in which $\hat{\Gamma}(x)$ is defined, rather than the function $\hat{\Gamma}(x)$ itself. To this end, $\hat{\Gamma}(x)$ is initially represented as a stochastic field which is distributed according to the \emph{prior} distribution
\begin{flalign}
\Pi[\hat{\Gamma}-\hat{\Gamma}_\mathrm{prior}] =
\frac{1}{\mathcal{N}}
\exp\left(
-\frac{1}{2}\left|
\hat{\Gamma}-\hat{\Gamma}_\mathrm{prior}
\right|^2_{\mathcal{K}_\mathrm{prior}}
\right)\;,
\end{flalign}
with
\begin{flalign}
&
\left|
\hat{\Gamma}
\right|^2_{\mathcal{K}_\mathrm{prior}} =
\int_0^\infty dx_1\int_0^\infty dx_2\,
\hat{\Gamma}(x_1)
\mathcal{K}^{-1}_\mathrm{prior}(x_1,x_2)
\hat{\Gamma}(x_2)\;,
\end{flalign}
and
\begin{flalign}
\mathcal{N}=\int \mathcal{D}\hat{\Gamma}\,
\Pi[\hat{\Gamma}]\;.
\end{flalign}
In the previous expressions $\mathcal{D}\hat{\Gamma}$ is the functional integration measure over the field $\hat{\Gamma}(x)$ while $\hat{\Gamma}_\mathrm{prior}(x)$ and $\mathcal{K}_\mathrm{prior}(x_1,x_2)$ are, respectively, the centre and the covariance of the prior distribution. By construction one thus has
\begin{flalign}
&
\hat{\Gamma}_\mathrm{prior}(x)
=
\int \mathcal{D}\hat{\Gamma}\, \Pi[\hat{\Gamma}-\hat{\Gamma}_\mathrm{prior}]\, \hat{\Gamma}(x)\;,
\nonumber \\[8pt]
&
\mathcal{K}_\mathrm{prior}(x_1,x_2)
=
\int \mathcal{D}\hat{\Gamma}\, \Pi[\hat{\Gamma}]\, \hat{\Gamma}(x_1)\, \hat{\Gamma}(x_2)\;,
\end{flalign}
and, in the following, we will only consider symmetric covariances, i.e.\
\begin{flalign}
\mathcal{K}_\mathrm{prior}(x_1,x_2)=\mathcal{K}_\mathrm{prior}(x_2,x_1)\;.
\end{flalign}

The goal now is to infer the \emph{posterior} distribution of $\hat{\Gamma}(x)$ at the generic point $x$ having observed (computed on the lattice) the values $\hat{\Gamma}_\mathrm{lat}^i$, with their (statistical and systematic) uncertainties, at the points $x^i_\mathrm{lat}$ for $i=0,\cdots,N_\mathrm{lat}-1$.

In order to introduce in the formulation the observational (lattice) errors, we consider the
stochastic vector $\vec \eta$, with elements $\eta^i=\eta(x^i_\mathrm{lat})$, distributed according to the multivariate Gaussian
\begin{flalign}
\mathbb{G}[\vec \eta;\hat \Sigma_\mathrm{lat}]
=\frac{1}{\sqrt{\det(2\pi\hat \Sigma_\mathrm{lat})}}
\exp\left(
-\frac{1}{2}\vec{\eta}^T \hat \Sigma_\mathrm{lat}^{-1}\vec{\eta}
\right)\;,
\end{flalign}
where $\hat \Sigma_\mathrm{lat}$ is the $N_\mathrm{lat}\times N_\mathrm{lat}$ covariance matrix of the non-perturbative lattice data (including both statistical and systematic errors). This allows us to connect the prior theoretical field $\hat{\Gamma}(x)$ to the lattice calculations by introducing the stochastic vector $\vec{\hat{\Gamma}_\mathrm{obs}}$ with elements
\begin{flalign}
\hat{\Gamma}_\mathrm{obs}^i= \hat{\Gamma}(x^i_\mathrm{lat}) + \eta^i\;.
\end{flalign}
The covariance of $\hat{\Gamma}_\mathrm{obs}^i$ and $\hat{\Gamma}_\mathrm{obs}^j$ is given by
\begin{flalign}
\hat \Sigma_\mathrm{prior}^{ij}+\hat \Sigma_\mathrm{lat}^{ij}
=
\int d\vec \eta\, \mathcal{D}\hat{\Gamma}\,
\hat{\Gamma}_\mathrm{obs}^i\, \hat{\Gamma}_\mathrm{obs}^j\,
\Pi[\hat{\Gamma}]\, \mathbb{G}[\vec \eta;\hat \Sigma_\mathrm{lat}]\;,
\end{flalign}
while the covariance of $\hat{\Gamma}_\mathrm{obs}^i$ and $\hat{\Gamma}(x)$ is given by
\begin{flalign}
\Delta^i
=
\int d\vec \eta\, \mathcal{D}\hat{\Gamma}\,
\hat{\Gamma}_\mathrm{obs}^i\, \hat{\Gamma}(x)\,
\Pi[\hat{\Gamma}]\, \mathbb{G}[\vec \eta;\hat \Sigma_\mathrm{lat}]\;,
\end{flalign}
where we called
\begin{flalign}
&
\hat \Sigma_\mathrm{prior}^{ij}
\equiv \mathcal{K}_\mathrm{prior}(x^i_\mathrm{lat},x^j_\mathrm{lat})\;,
\nonumber \\[8pt]
&
\Delta^i\equiv \mathcal{K}_\mathrm{prior}(x,x^i_\mathrm{lat})\;.
\end{flalign}
These are the relevant elements of the prior covariance $\mathcal{K}_\mathrm{prior}$
that we conveniently collect into the matrix $\hat \Sigma_\mathrm{prior}$ and the vector $\vec \Delta$.

By introducing the total covariance matrix
\begin{flalign}
\hat \Sigma_\mathrm{tot}
=
\left(
\begin{array}{cc}
\mathcal{K}_\mathrm{prior}(x,x) & \vec \Delta^T
\\
\vec \Delta & \hat \Sigma_\mathrm{prior}+\hat \Sigma_\mathrm{lat}
\end{array}
\right)\;,
\end{flalign}
we now have the joint distribution of $\hat{\Gamma}(x)$ and $\vec{\hat{\Gamma}_\mathrm{obs}}$, which is given by the multivariate Gaussian
\begin{flalign}
\mathbb{G}
\left[\left(\hat{\Gamma}(x)-\hat{\Gamma}_\mathrm{prior}(x),\vec{\hat{\Gamma}_\mathrm{obs}}-\vec{ \hat{\Gamma}_\mathrm{prior}} \right);\hat \Sigma_\mathrm{tot}
\right]\;.
\end{flalign}
This, by using the formulae of Appendix~A of Ref.~\cite{DelDebbio:2024lwm}, can be conditioned as follows,
\begin{flalign}
&
\mathbb{G}
\left[\left(\hat{\Gamma}(x)-\hat{\Gamma}_\mathrm{prior}(x),\vec{\hat{\Gamma}_\mathrm{obs}}-\vec{ \hat{\Gamma}_\mathrm{prior}} \right);\hat \Sigma_\mathrm{tot}
\right] =
\nonumber \\[8pt]
&
\qquad
\mathbb{G}
\left[\hat{\Gamma}(x)-\hat{\Gamma}_\mathrm{post}(x); \Delta_\mathrm{post}(x)
\right]
\times
\nonumber \\[4pt]
&
\qquad
\qquad
\times
\mathbb{G}
\left[\vec{\hat{\Gamma}_\mathrm{obs}}-\vec{ \hat{\Gamma}_\mathrm{prior}};\hat \Sigma_\mathrm{prior}+\hat \Sigma_\mathrm{lat}
\right]\;,
\end{flalign}
where $\mathbb{G}\left[\hat{\Gamma}(x)-\hat{\Gamma}_\mathrm{post}(x); \Delta_\mathrm{post}(x)\right]$ is the posterior distribution of $\hat{\Gamma}(x)$ given the observations $\vec{\hat{\Gamma}_\mathrm{obs}}$. The central value and the standard deviation of the posterior distribution are given by
\begin{flalign}
&
\hat{\Gamma}_\mathrm{post}(x)
=
\hat{\Gamma}_\mathrm{prior}(x)
+
\vec \Delta^T\frac{1}{\hat \Sigma_\mathrm{prior}+\hat \Sigma_\mathrm{lat}}
\left( \vec{\hat{\Gamma}_\mathrm{obs}}-\vec{ \hat{\Gamma}_\mathrm{prior}}\right),
\nonumber \\[8pt]
&
\Delta_\mathrm{post}(x)
=
\mathcal{K}_\mathrm{prior}(x,x)
-
\vec \Delta^T\frac{1}{\hat \Sigma_\mathrm{prior}+\hat \Sigma_\mathrm{lat}}
\vec \Delta\;.
\label{eq:bayesianres}
\end{flalign}
By using now the fact that by performing our lattice calculation we observed
\begin{flalign}
\vec{\hat{\Gamma}_\mathrm{obs}} = \vec{ \hat{\Gamma}_\mathrm{lat}}\,,
\end{flalign}
\cref{eq:bayesianres} can be used to predict $\hat{\Gamma}(x)$ and its error once a choice for $\hat{\Gamma}_\mathrm{prior}(x)$ and $\mathcal{K}_\mathrm{prior}(x_1,x_2)$ has been made.

Also in this case there is a \emph{very important} remark that needs to be done. When, in order to get a physical result, an extrapolation/interpolation is necessary, it is
unavoidable to assume that the knowledge of a quantity in a region (in our case the knowledge of the discrete data $\hat{\Gamma}(x^i_\mathrm{lat})$ and their lattice covariance) provides useful information on that quantity in a close-by region. This in fact means that it is unavoidable to assume some degree of regularity of the function/field that has to be inferred (in our case $\hat{\Gamma}(x)$). Within the fitting-function approach this assumption, hopefully motivated by the underlying theory, allows to perform the crucial step, i.e.\ the choice of the fitting function (in our case $\hat{\Gamma}^{(n)}(x)$ of \cref{eq:freqfitfunc}). The fitting function has to be sufficiently general to allow any possible (theory motivated) behaviour of the quantity outside the explored region and, at the same time, sufficiently regular to allow the extrapolation/interpolation. In our case, as discussed in the previous subsection, the rational non-perturbative contribution $\hat{\Gamma}_\mathrm{np}^{(n)}(x)$ to $\hat{\Gamma}^{(n)}(x)$ meets both these requirements.

In the functional-space approach the crucial step is the choice of the covariance $\mathcal{K}_\mathrm{prior}(x_1,x_2)$. Less important, although it allows to leverage on general properties of the quantity and to impose possible exact constraints coming from theory, is the choice of the prior model $\hat{\Gamma}_\mathrm{prior}(x)$. As pointed out in Ref.~\cite{DelDebbio:2024lwm}, in the extreme situation in which the quantity under investigation is a distribution it is impossible to extrapolate the knowledge gained at some point $x$ even at the infinitely close-by point $x+0^+$: in this case the covariance is ultra-local, i.e.\ $\mathcal{K}_\mathrm{prior}(x_1,x_2)\propto \delta(x_1-x_2)$. The space of ``extrapolatable'' functions can be very well parametrized by considering covariances that are proportional to Gaussians\footnote{See~\cite{10.1093/gji/ggz520} and references therein for more details concerning this point: a more general parametrization can be obtained in terms of the so-called Mat\'ern function, depending on a parameter $\nu$ which can be put in correspondence with the degree of differentiability of the functions. In the $\nu\mapsto \infty$ limit the Mat\'ern function becomes a Gaussian.}, i.e. $\mathcal{K}_\mathrm{prior}(x_1,x_2)\propto \exp(-(x_1-x_2)^2/(2\ell^2))$, with the parameter $\ell$ playing the role of the correlation length. In this way, in the $\ell\mapsto 0$ limit one recovers the distributional case while, at finite $\ell$, observations performed at $x\pm\ell$ provide useful information on the behaviour of the function at $x$.

\begin{figure}
\includegraphics[width=\linewidth]{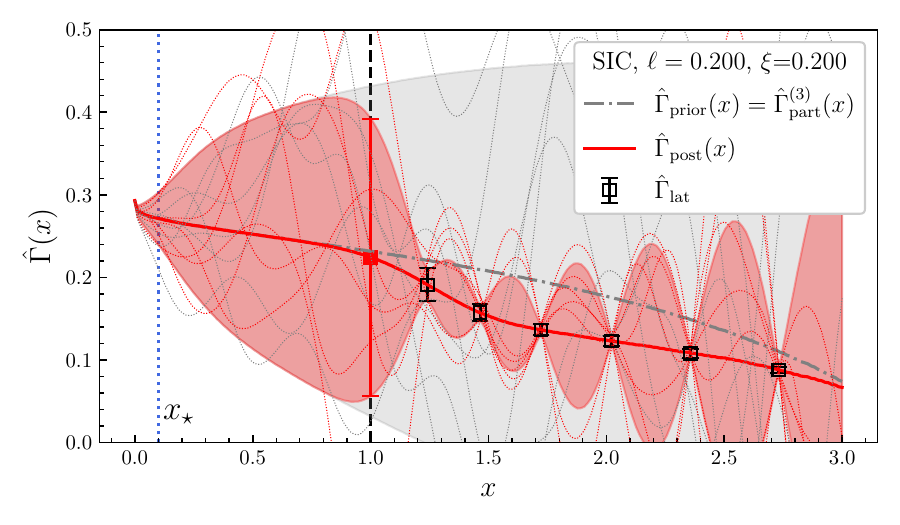}
\includegraphics[width=\linewidth]{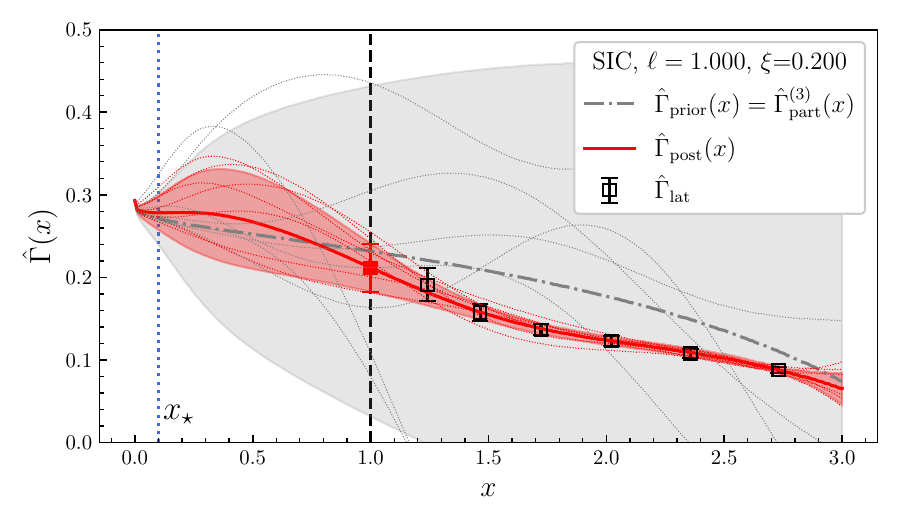}
\includegraphics[width=\linewidth]{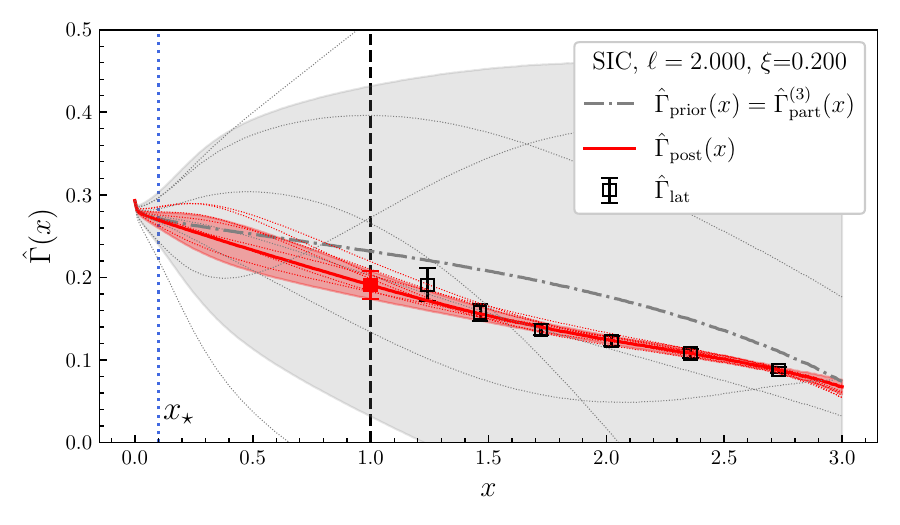}
\caption{Results of the interpolation to $x=1$ employing the functional-space approach of \cref{sec:bayesian} for different values of the correlation length parameter $\ell$. The other parameters of the covariance of the prior distribution are fixed at $x_\star=0.1$, $\xi=0.2$ and $\varepsilon=0.1$. The gray and red bands represent, respectively, the prior and posterior distribution. The light grey curves represent instances of $\hat \Gamma(x)$ drawn from the prior distribution. Analogously, the light red curves represent instances of $\hat \Gamma(x)$ drawn from the posterior distribution.}
\label{fig:bayesian_fits}
\end{figure}

In our analysis, for the central value of the prior distribution we considered again the partonic result,
\begin{flalign}
 \hat{\Gamma}_\mathrm{prior}(x)=\hat{\Gamma}_\mathrm{part}^{(n)}(x),
\end{flalign}
with $\mu_s(x)=M_{\bar B_s}/{2x}$ and for different values of $n$. Concerning the prior covariance, we adopted the following parametrization
\begin{flalign}
\mathcal{K}_\mathrm{prior}(x_1,x_2)=\xi(x_1)\xi(x_2)\,\exp\left(-\frac{(x_1-x_2)^2}{2\ell^2}\right) \;,
\label{eq:Kprior}
\end{flalign}
with
\begin{flalign}
\xi(x)=\frac{\xi \sqrt{x}}{1+e^{-(x-x_\star)/\varepsilon}}\;.
\end{flalign}
With this choice we have that
\begin{flalign}
&
\mathcal{K}_\mathrm{prior}(x,x)=\xi^2x, \qquad x\gg x_\star\,;
\nonumber \\[8pt]
&\mathcal{K}_\mathrm{prior}(x,x)=0, \qquad x\ll x_\star.
\end{flalign}
The parameter $x_\star$ represents a threshold value below which the posterior distribution is forced to rapidly become very narrow around the central value of the prior distribution, i.e. $\hat{\Gamma}_\mathrm{part}^{(n)}(x)$. This is a desired feature because asymptotic freedom implies that the partonic result has to be a very good approximation for very small values of $x$.
Instead, for $x\gg x_\star$ the prior covariance grows linearly, with an amplitude given by the square of the parameter $\xi$. This is also a desirable feature because, according to the OPE results, the leading non-perturbative correction to the partonic result is $O(x)$. The switching parameter $\varepsilon$ allows a smooth transition between the $x\ll x_\star$ and $x\gg x_\star$ regimes. The amplitude parameter $\xi$ must be large enough, so that, given also the choices of $x_\star$ and $\varepsilon$, the prior covariance in the region $x\ge 1$ should not pose a limiting constraint for the posterior covariance. In summary, the tuning of the parameters $x_\star$, $\epsilon$ and $\xi$ is an easy and not particularly delicate task.

Conversely, as already emphasized, the crucial and delicate task is that of the tuning of the correlation length parameter $\ell$: the smaller is the value of $\ell$, the larger is the error on the interpolated result and vice-versa. Moreover, the larger $\ell$ is, the stronger is the impact of the theory constraint $\hat{\Gamma}(0)=\hat{\Gamma}_\mathrm{part}^{(0)}(0)$ on the interpolation at $\hat{\Gamma}(1)$.

\cref{fig:bayesian_fits} shows three interpolations, performed at fixed $x_\star=0.1$, $\xi=0.2$, $\varepsilon=0.1$
and for increasing values of the correlation length parameter $\ell$. In all panels, the grey band is the prior covariance ($\mathcal{K}_\mathrm{prior}(x,x)$), the red band is the posterior error $\Delta_\mathrm{post}(x)$, the solid dash-dotted grey curve is $\hat{\Gamma}_\mathrm{prior}(x)$, which in this case is $\hat{\Gamma}^{(3)}_\mathrm{part}(x)$, and the solid red curve is $\hat{\Gamma}_\mathrm{post}(x)$. The light grey/red curves are possible instances of $\hat \Gamma(x)$ drawn from the prior/posterior distribution. The black points are the lattice data, the ones corresponding to the SIC analysis order, while the red point is the result of the interpolation, $\hat{\Gamma}_\mathrm{post}(1)$ and its error $\Delta_\mathrm{post}(1)$, at the physical point $x=1$ marked by the dashed black vertical line. The dotted blue vertical line marks our choice of the parameter $x_\star$.

In the top panel, in order to illustrate the possible implications associated with the choice of the correlation length parameter $\ell$, we show a rather unphysical (or at least borderline) situation. In this case $\ell=0.2$ and we are now going to argue that this value is too small. This is already evident by looking at the possible instances of the field $\hat \Gamma(x)$ drawn from the prior (light grey curves) and posterior (light red curves) distributions: with such a small value of $\ell$ these start to exhibit a distributional character. Actually, from our lattice exploration of the region $1.2< x < 2.8$ we know that it is very reasonable to assume that $\hat{\Gamma}(x)$ is very smooth on a length scale $\ell \sim 2$. Moreover, from our knowledge of the partonic result we can also reasonably assume that $\hat{\Gamma}(x)$ is quite smooth outside the lattice data region, except perhaps for very small values of $x$ where the logarithms present in $\hat{\Gamma}_\mathrm{part}^{(3)}(x)$ generate the rather steep behaviour in approaching the static limit $\hat{\Gamma}_\mathrm{part}^{(0)}(0)$. This, however, is an exact theoretical constraint that we take into account with a reasonable choice of $x_\star$, the parameter that effectively separates the ``fully perturbative'' region from the non-perturbative one. The effect of a too small value of $\ell$ are the bubbles of the posterior covariance in the middle of the lattice points and the fact that for $x<1$ the posterior is only constrained by the prior, neither by the data, nor by the theory constraint at $x=0$. On the one hand, we cannot definitely exclude that this is a possible behaviour of $\hat{\Gamma}(x)$. On the other hand, we are confident that any reasonable data-analyst would consider $\ell=0.2$ too small a value in this context.

The phenomenology rapidly changes in passing from the top to the bottom panel of \cref{fig:bayesian_fits} where $\ell=2$. In this case the result is starting to get more similar to the one that we got by using the fitting-function approach (see \cref{fig:padeperturbativeorders}). As it can be seen, the possible instances of $\hat \Gamma(x)$ are much smoother but, at same time, the amplitude of the prior covariance allows to explore sufficiently general possible behaviours of $\hat \Gamma(x)$. Moreover, the prior distribution is sufficiently wide not to have any sizeable effect on the posterior result in the region $x\ge 1$. On the other hand, the error on the extrapolated point is considerably larger than the one that we get in the fitting-function approach. This is due to the fact that, with this value of the correlation length, the precise lattice points at $x\sim 2$ and the exact theory result at $x=0$ do have a non-negligible impact on the interpolation, but one that is not as strong as in the fitting-function case. An intermediate situation is shown in the middle panel.

\begin{figure}
\includegraphics[width=\linewidth]{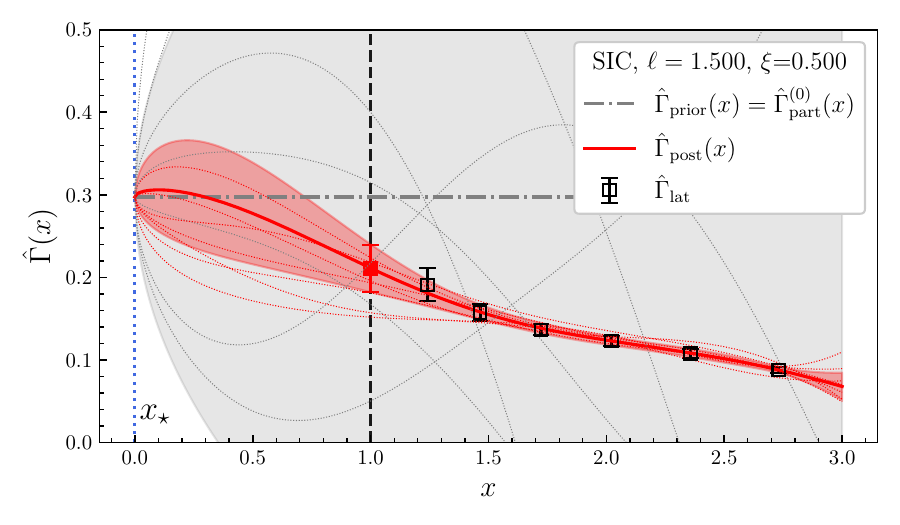}
\includegraphics[width=\linewidth]{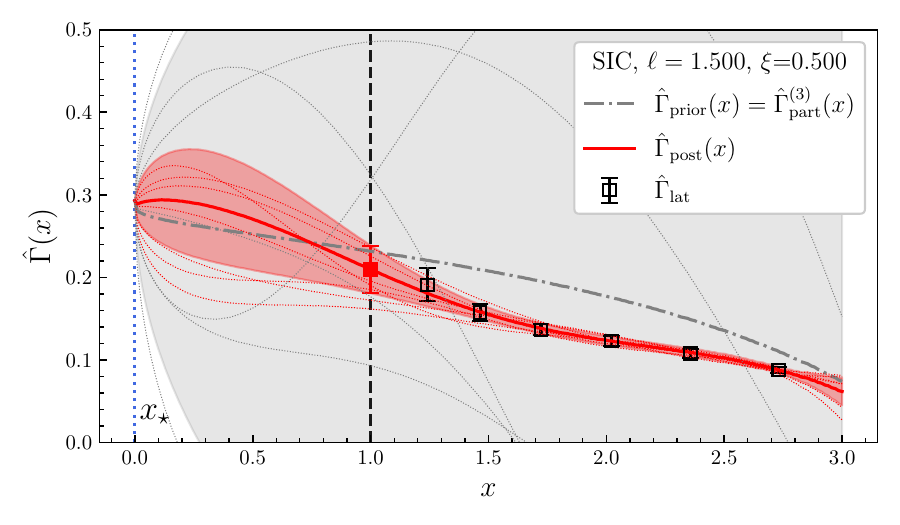}
\caption{Same as \cref{fig:bayesian_fits} but with $\xi=0.5$, $x_\star=\varepsilon=0$ and $\ell=1.5$. The top panel corresponds to the choice $\hat \Gamma_\mathrm{prior}(x)=\hat \Gamma_\mathrm{part}^{(0)}(x)$ while the bottom panel to the choice $\hat \Gamma_\mathrm{prior}(x)=\hat \Gamma_\mathrm{part}^{(3)}(x)$.}
\label{fig:bayesian_fits2}
\end{figure}

Two more examples are illustrated in \cref{fig:bayesian_fits2}. These show explicitly that, if the prior distribution is sufficiently wide, $\xi=0.5$ in this case, the choice of its central value $\hat \Gamma_\mathrm{prior}(x)$ is irrelevant. Moreover, these two examples show what happens if the fully perturbative region is reduced to the static point only, $x_\star=\varepsilon=0$, and with the choice $\ell=1.5$ of the correlation length parameter.

\begin{figure}
    \centering
    \includegraphics[width=\linewidth]{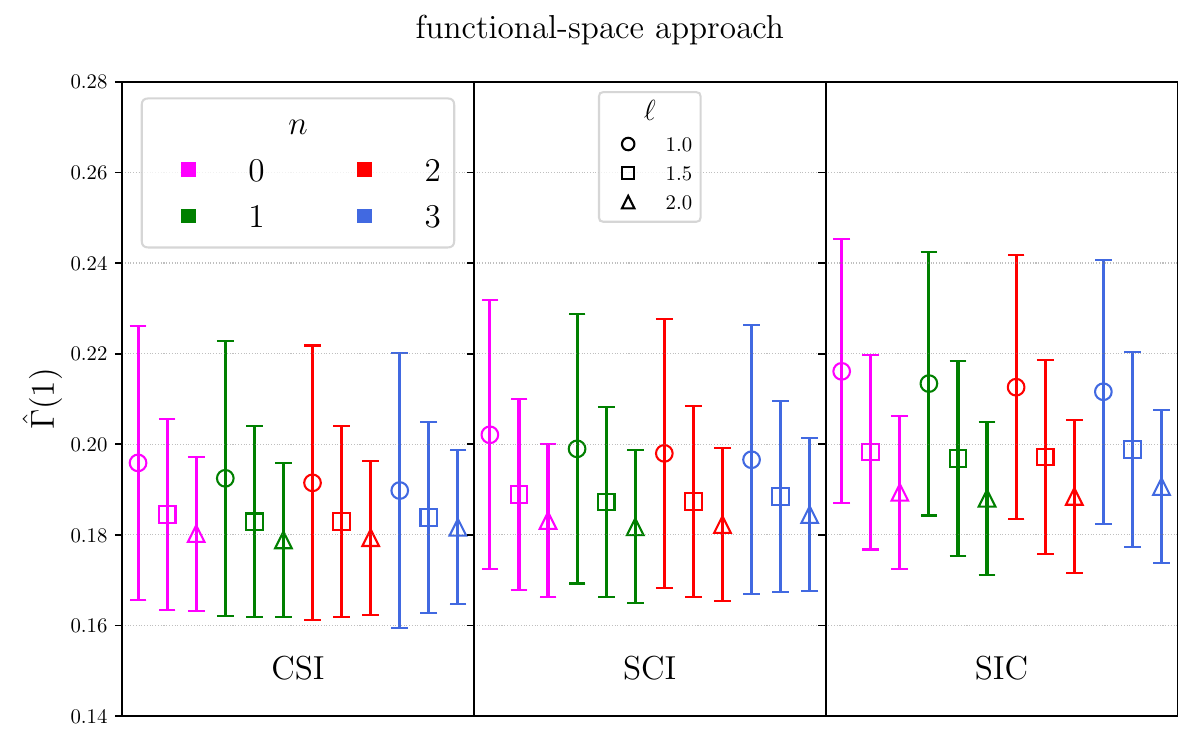}
    \caption{The result of the interpolation of $\hat{\Gamma}$ at the physical point by using the functional-space approach with $x_\star=0.1$, $\varepsilon=0.1$ and $\xi=0.2$ for different values of the correlation length parameter $\ell$ and different choices of the prior model $\hat{\Gamma}_\mathrm{prior}(x)=\hat{\Gamma}_\mathrm{part}^{(n)}(x)$ with $\mu_s(x)= \frac{M_{\bar B_s}}{2x}$.
    In each panel, magenta points corresponds to $n=0$, green points to $n=1$, red points to $n=2$ and the blue points to $n=3$. Circles correspond to $\ell=1.0$, squares to $\ell=1.5$ and triangles to $\ell=2$.
    The left, center and right panels correspond to the analysis orders CSI, SCI and SIC, respectively.}
    \label{fig:comparisonbayesian0}
\end{figure}

In \cref{fig:comparisonbayesian0} we compare the results of the functional-space interpolations obtained in the different analysis orders with $x_\star=0.1$, $\varepsilon=0.1$ and  $\xi=0.2$, i.e.\ with a sufficiently wide prior distribution, for different values of $\ell$ and different choices of the prior model. This is the analogue of \cref{fig:frequentistcomparison} in the fitting-function approach and, as it can be seen, the errors that we get in the functional-space case are considerably larger. In the previous discussion we explained that smaller errors can be obtained by considering larger values of $\ell$ and/or smaller values of $\xi$.

\begin{figure}
\includegraphics[width=\linewidth]{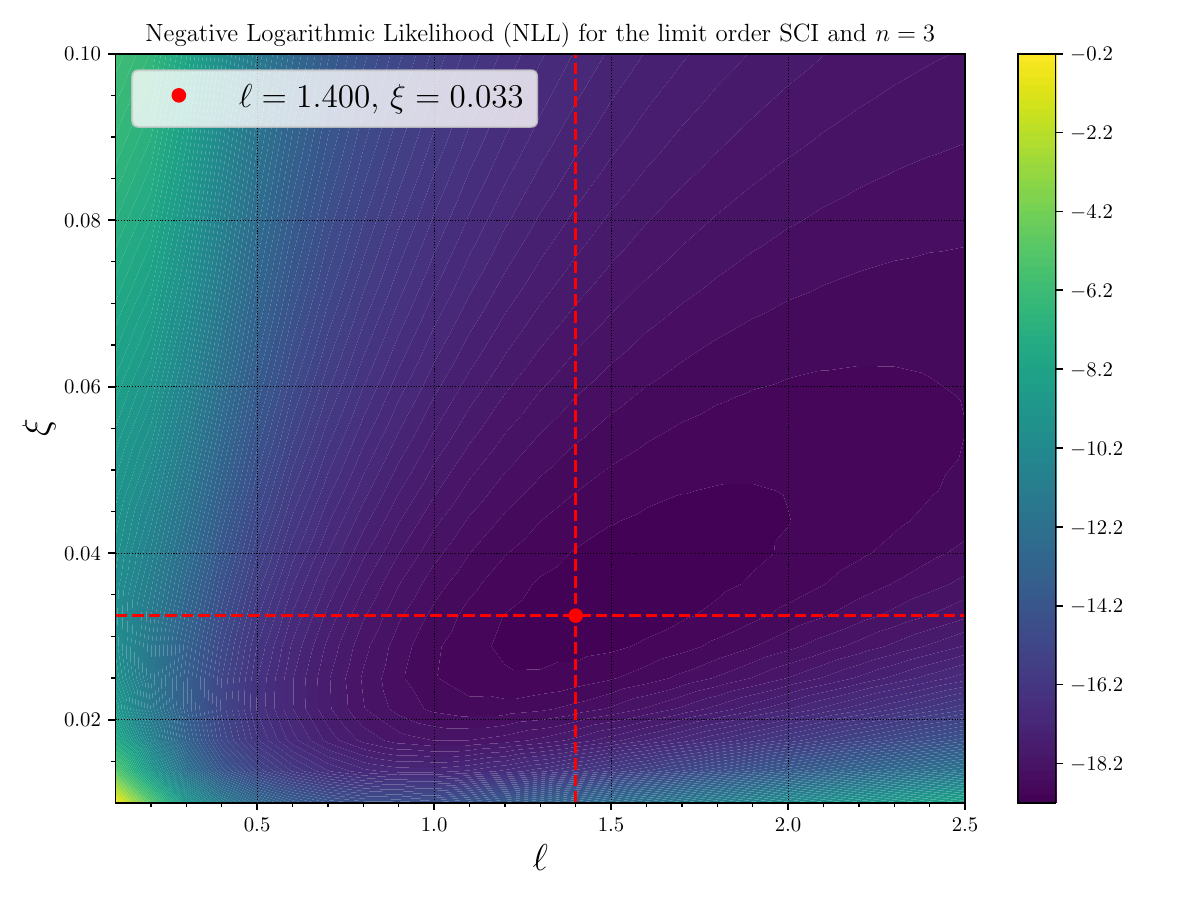}
\includegraphics[width=\linewidth]{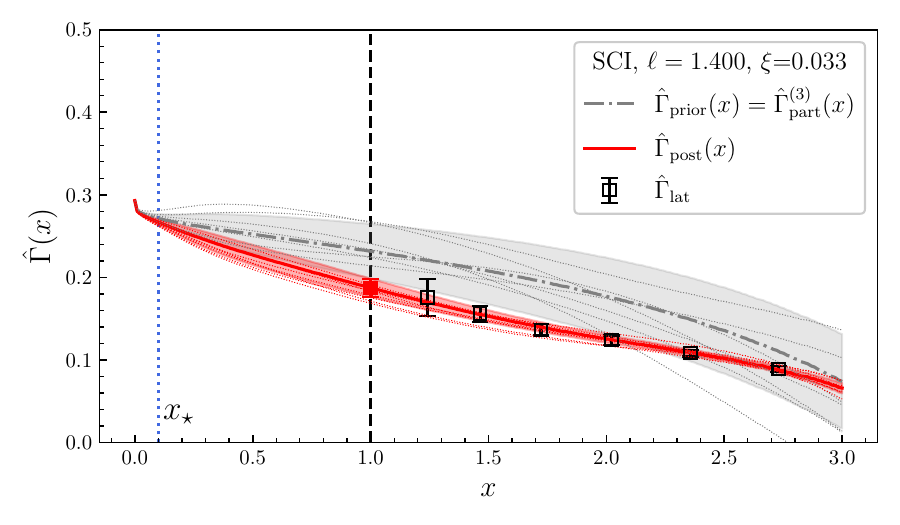}
\caption{\emph{Top-plot}: heat map of the NLL in case of the analysis order SIC and $n=3$ at fixed $x_\star=0.1$ and $\varepsilon=0.1$. The red point marks the minimum and it is obtained after a scan in a $25\times 25$ grid covering the interval $\xi=[0.01,0.1]$ and $\ell=[0.1,2.5]$. \emph{Bottom-plot}: the same as Figure~\ref{fig:bayesian_fits} but in case of the analysis order SCI and with $\ell$ and $\xi$ obtained from the minimization of the NLL.}
\label{fig:NLL}
\end{figure}

\begin{figure}
\includegraphics[width=\linewidth]{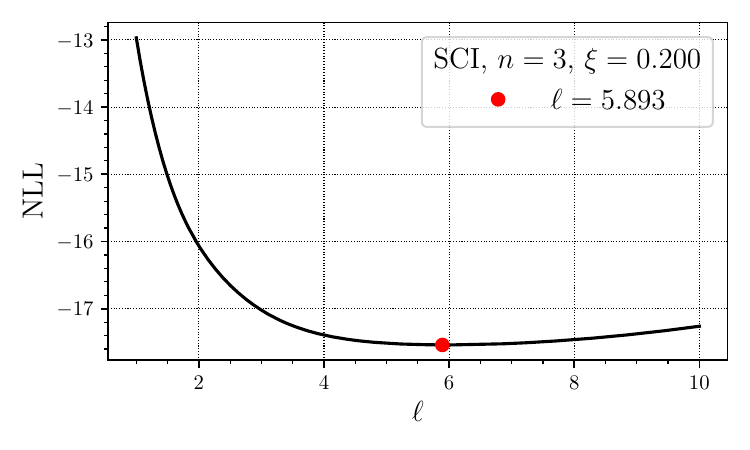}
\includegraphics[width=\linewidth]{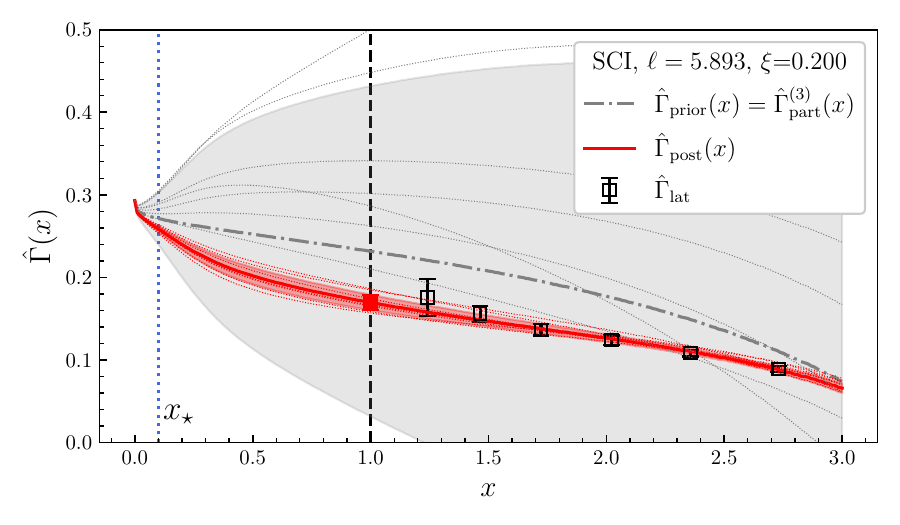}
\caption{\emph{Top-plot}: NLL as function of $\ell$ at fixed $\xi=0.2$ for the limits' order SCI and $n=3$. The red point marks the minimum value. \emph{Bottom-plot}: corresponding plot of the prior and posterior distributions.}
\label{fig:NLL_fixedxi}
\end{figure}

From a fully Bayesian perspective, the ``optimal'' values of the parameters of the prior distribution have to be obtained by maximizing the probability of observing the data. By taking this perspective, we determined the Bayesian-optimal values of $\xi$ and $\ell$ by minimizing the Negative Logarithmic Likelihood (NLL)
\begin{flalign}
-\log\left( \mathbb{G}
\left[\vec{\hat{\Gamma}_\mathrm{lat}}-\vec{ \hat{\Gamma}_\mathrm{prior}};\hat \Sigma_\mathrm{prior}+\hat \Sigma_\mathrm{lat}
\right]\right)\;.
\label{eq:NNL}
\end{flalign}
The result, in the case of the SCI analysis order and $n=3$, is shown in \cref{fig:NLL}. As it can be seen, by taking this perspective, the prior distributions considered in \cref{fig:bayesian_fits,fig:bayesian_fits2} were too wide for the choice of the prior central value $\hat{\Gamma}_\mathrm{part}^{(3)}(x)$. Indeed the resulting optimal value of the amplitude parameter is $\xi\sim 0.03$, while the optimal value of the correlation length parameter, $\ell=1.4$, is consistent with the ones that we considered very reasonable choices in \cref{fig:comparisonbayesian0}. To close this discussion, we show in \cref{fig:NLL_fixedxi} what happens if we insist in choosing a rather wide prior distribution (we set $\xi=0.2$) and minimize the NLL w.r.t. $\ell$ only. The resulting optimal value of the correlation length parameter is larger ($\ell=5.9$) but the result of the interpolation is fully compatible with the one obtained in \cref{fig:NLL} with an error of the same size.

\begin{figure}
\centering
\includegraphics[width=\linewidth]{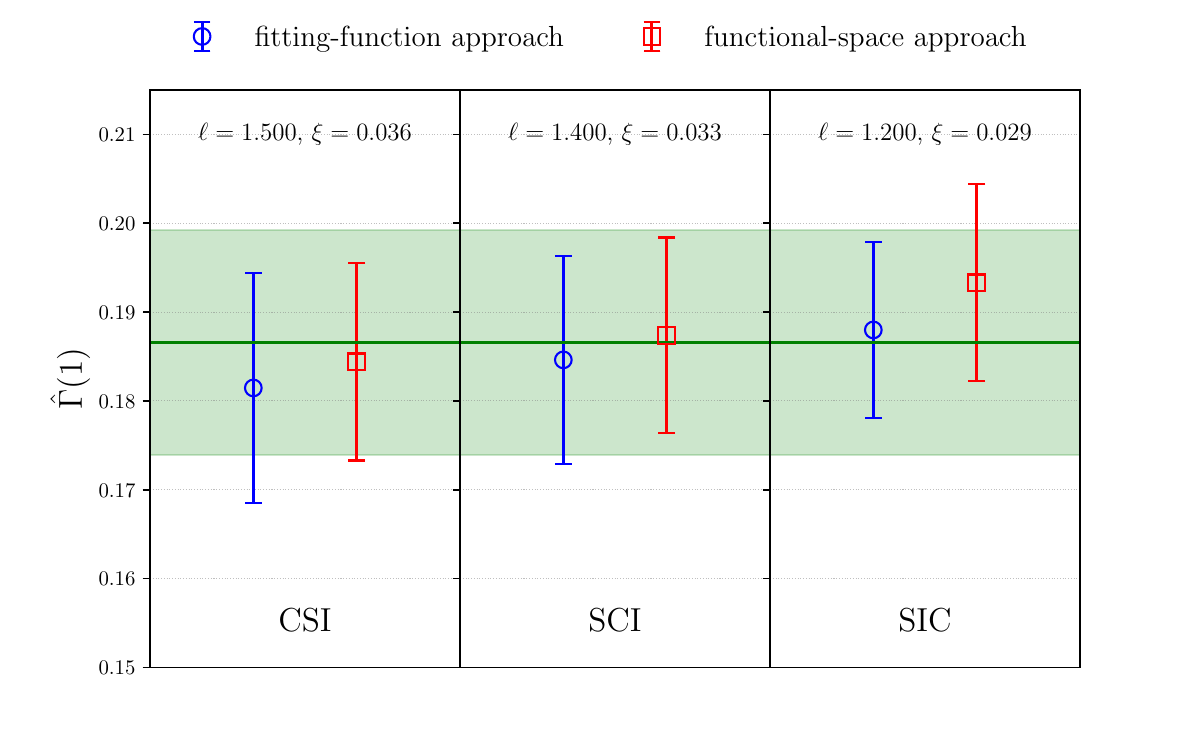}
\caption{Comparison of  $\hat{\Gamma}$ at the physical point between the fitting-function approach (with $\mu_s(x)=\frac{1}{2}\frac{M_{\bar B_s}}{x}$ ) and the functional-space approach where the parameters $\ell$ and $\xi$ are obtained from the minimization of the NLL. In both cases $n=3$.   The left, center and right panels correspond to the analysis orders CSI, SCI and SIC, respectively. The green band corresponds to our final determination, $\hat{\Gamma}(1)=0.187(10)(8)[13]$, whose central value is obtained from the average of the six points and the error is the combination in quadrature of the statistical and systematic uncertainties. The systematic error is estimated according to the formulas presented in  \cref{sec:resultswithhlt}.}
\label{fig:final_comparison_lattice}
\end{figure}

In \cref{fig:final_comparison_lattice} we show the comparison, in the different analysis orders, of the results obtained in the functional-space approach by taking the Bayesian-optimal values of the parameters $\xi$ and $\ell$ (red points), with the results obtained in the fitting-function approach (blue points). We consider it very reassuring that the errors of the two approaches are now of the same order of magnitude and that all the determinations are  well compatible with each other.

To quote our current result for $\Gamma$ at the physical point we take the central value and the statistical error from the correlated average of the six determinations in \cref{fig:final_comparison_lattice} and estimate the systematic error associated with the interpolation by using \cref{eq:pull,eq:systematic_error}. This (the green band) gives
\begin{flalign}
\hat{\Gamma}(1)=0.187(10)(8)[13]\;.
\label{eq:finalhat}
\end{flalign}
The systematic error can be split into several parts, 68\% of the systematic error being from the spread between the different analysis orders, and 32\% from the systematic errors introduced by the HLT reconstruction, the continuum and $\sigma \mapsto 0$ limit and the integration. The latter part of the systematic error is dominated by the HLT uncertainty.
Notice that, by taking the spread of the results associated with the different analysis orders into account, we further enlarge the systematic errors associated with the continuum limit, the $\sigma\mapsto 0$ extrapolation and the numerical integration. We do this to account for the systematics associated with the infinite-volume extrapolation which, presently, we do not estimate from our data. Actually, the $\bar B_s\mapsto X_{\bar s c} \ell \bar \nu$ decay rate is expected to be much less sensitive to long-distance effects than the $D_s\mapsto X \ell \bar \nu$ one, and in Refs.~\cite{DeSantis:2025yfm,DeSantis:2025qbb} we found that the FVE systematic errors were below the 1\% level, and therefore negligible at the current level of precision. As already said, we will explicitly check the validity of this very reasonable assumption by using the B48 and B96 ensembles of \cref{tab:iso_EDI_FLAG} in our forthcoming publication.

Multiplying $\hat{\Gamma}(1)$ of \cref{eq:finalhat} by $\bar{\Gamma}(1)=1.325\times 10^{-10}$~GeV, we get our current result,
\begin{flalign}
\Gamma(1)=2.47(13)(10)[17]\times 10^{-11}\;\mathrm{GeV}\;.
\end{flalign}

\section{
\label{sec:conclusion}
Conclusions and Outlooks
}

\begin{figure}
\centering
\includegraphics[width=\linewidth]{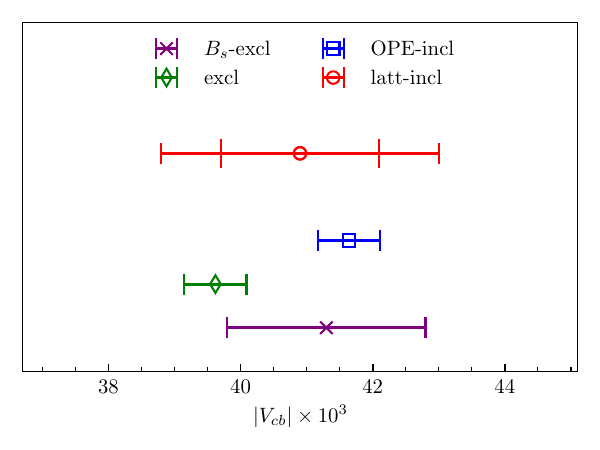}
\caption{Comparison between our lattice determination (red), the exclusive determination (green), the OPE-inclusive one (blue), and the determination from the exclusive $\bar{B}_s\mapsto D_s^{(\star)}\ell \bar{\nu}$ decay (purple) \cite{Martinelli:2022xir} of the $|V_{cb}|$ matrix element. The smaller red bar is our theoretical error while the larger bar takes the experimental error into account.}
\label{fig:final_comparison}
\end{figure}

In this paper we presented, and thoroughly discussed from both the theoretical and numerical viewpoints, an effective strategy for the non-perturbative calculation of the inclusive semileptonic decay rate $\Gamma$ of the process $\bar B_s\mapsto X_{\bar s c} \ell \bar \nu$. We also obtained a first result which, in the near future, we are going to improve by repeating the analysis performed here using more lattice gauge ensembles.

We regard the theoretical and methodological results obtained in this paper as particularly important.

We have introduced a new numerical strategy for the computation of the required four-point lattice correlators. This allowed us to solve the problem, encountered with the previous state-of-the-art strategy~\cite{Gambino:2020crt,Gambino:2022dvu,Barone:2023tbl,DeSantis:2025yfm,DeSantis:2025qbb,Kellermann:2026sgp}, of the rapid deterioration of the signal-to-noise ratio with the increasing of the meson masses.

We also devised an interpolation strategy which allows to obtain the physical decay rate by combining non-perturbative lattice results at lighter than physical heavy meson masses  with the OPE analytical predictions, which become exact in the static limit. By using both a fitting-function and a functional-space approach, we presented a very detailed discussion of this crucial analysis step and showed the effectiveness of the proposed interpolation strategy.

By using three ETMC ensembles at physical quark masses, with $a\ge 0.056$~fm and $L\simeq 5.5$~fm, we obtained our current theoretical prediction\footnote{We remind the reader that the definition of $\Gamma$ used in the main body of the paper does not include the $\vert V_{cb}\vert^2$ factor.}
\begin{flalign}
\frac{\Gamma^\mathrm{latt-incl}[\bar B_s\mapsto X_{\bar s c} \ell \bar \nu]}{\vert V_{cb}\vert^2}
=2.47(13)(10)[17]\times 10^{-11}\;\mathrm{GeV}\;.
\end{flalign}
The total error, in square brackets, includes the systematics associated with the spectral reconstruction, the continuum extrapolations, the numerical integration and the interpolation to the physical point. On the one hand, our current estimate of the error has to be considered rather conservative. It has been generously estimated by performing the various steps of the analysis with different methods and in different orders. On the other hand, presently, our error does not include a data-driven estimate of finite-volume effects; however, from our previous computation of the $D_s\mapsto X\ell \bar{\nu}$ decay rate, we expect these to be negligible at the current level of precision.

Presently, an experimental measurement of the $\Gamma[\bar B_s\mapsto X_{\bar s c} \ell \bar \nu]$ decay rate is not available. Nevertheless, we can consider the measurement of $\Gamma[\bar B_s\mapsto X \ell \bar \nu]$, which includes the contribution of the $X_{\bar s u}$ flavour channel, performed in Ref.~\cite{Belle:2012mwf} (see also~\cite{BaBar:2011sxq}),
\begin{flalign}
\Gamma^\mathrm{Belle}[\bar B_s\mapsto X \ell \bar \nu]
=4.14(35)\times 10^{-14}\;\mathrm{GeV}\;.
\end{flalign}
Indeed, under the theory motivated assumption\footnote{The assumption is supported by the first-principles analysis of the different contributions in the $D_s$ case performed in Refs.~\cite{DeSantis:2025yfm,DeSantis:2025qbb} and, by assuming U-spin symmetry, from the results $\mathrm{Br}[\bar B\mapsto X_{\bar d c} \ell \bar \nu]=10.63(15)\times 10^{-2}$ and $\mathrm{Br}[\bar B\mapsto X_{\bar d u} \ell \bar \nu]=1.92(21)\times 10^{-3}$ quoted in~\cite{HeavyFlavorAveragingGroupHFLAV:2024ctg}.} that $\Gamma[\bar B_s\mapsto X_{\bar s u} \ell \bar \nu]/|V_{ub}|^2$ is of the same order of magnitude as $\Gamma[\bar B_s\mapsto X_{\bar s c} \ell \bar \nu]/|V_{cb}|^2$, and by taking into account the $|V_{ub}|^2/|V_{cb}|^2\simeq 0.008$ suppression factor, it is reasonable to neglect the $X_{\bar s u}$ contribution at the present level of accuracy. Under this assumption, we get
\begin{flalign}
\vert V_{cb}\vert^\mathrm{latt-incl} = 40.9(1.2)^\mathrm{latt}(1.7)^\mathrm{exp}[2.1]^\mathrm{tot} \times 10^{-3}\;.
\label{eq:ourvcb}
\end{flalign}
This can be compared with the current best determinations coming from exclusive decays~\cite{HeavyFlavorAveragingGroupHFLAV:2024ctg}~(see also Refs.~\cite{Martinelli:2023fwm,Martinelli:2024bov,Bordone:2024weh})
\begin{flalign}
\vert V_{cb}\vert^\mathrm{excl} = 39.62(47)\times 10^{-3}\;,
\end{flalign}
and, by using the OPE with phenomenological inputs on the theory side, from inclusive decays~\cite{Carvunis:2025vab}
\begin{flalign}
\vert V_{cb}\vert^\mathrm{OPE-incl} = 41.64(47)\times 10^{-3}\;.
\end{flalign}
The comparison is shown in \cref{fig:final_comparison} where we also show the result coming from exclusive $B_s$ decays obtained in~\cite{Martinelli:2022xir}.
With our current error, which is experimentally dominated, we cannot solve this long-standing and crucially important puzzle. On the other hand, our current result further motivates us to complete this challenging and very interesting project.

In the (near) future, we plan to add an ensemble at a finer value of the lattice spacing ($a\simeq 0.05$~fm) and two more ensembles at different physical volumes as well as to considerably improve the statistics of all ensembles. This will allow us to have much better control on the continuum extrapolations, a crucially important step in any $b$-physics lattice calculation, and to numerically perform the infinite volume limit. Given the results obtained here, we are fully confident that we will be able to reduce the total final error well below the current 7\% level of accuracy.

As a further step, we will compute on the same gauge ensembles the exclusive decay rate for the process $\bar B_s\mapsto D_s \ell \bar \nu$ and perform a correlated extraction of $V_{cb}$ from both the inclusive and exclusive channels. We will then certainly extend our study to the case of $B$ decays.

\section{Acknowledgements}
The authors gratefully acknowledge the Gauss Centre for Supercomputing e.V. (www.gauss-centre.eu) for funding this project by providing computing time on the GCS Supercomputer JUWELS~\cite{JUWELS} at J\"ulich Supercomputing Centre (JSC) and on the GCS Supercomputers SuperMUC-NG at Leibniz Supercomputing Centre.
The authors acknowledge the Texas Advanced Computing Center (TACC) at The University of Texas at Austin for providing HPC resources (Project ID PHY21001). The authors gratefully acknowledge PRACE for awarding access to HAWK at HLRS within the project with Id Acid 4886. We acknowledge the Swiss National Supercomputing Centre (CSCS) and the EuroHPC Joint Undertaking for awarding this project access to the LUMI supercomputer, owned by the EuroHPC Joint Undertaking, hosted by CSC (Finland) and the LUMI consortium through the Chronos programme under project IDs CH17-CSCS-CYP. We acknowledge EuroHPC Joint Undertaking for awarding the project ID EHPC-EXT-2023E02-052 access to MareNostrum5 hosted by the Barcelona Supercomputing Center, Spain.

This work has been supported  by the MKW NRW under the funding code NW21-024-A as part of NRW-FAIR and by the Italian Ministry of University and Research (MUR) and the European Union (EU) – Next Generation EU, Mission 4, Component 1, PRIN 2022, CUP F53D23001480006 and CUP D53D23002830006.
We acknowledge support from the ENP, LQCD123, SFT, and SPIF Scientific Initiatives of the Italian Nuclear Physics Institute (INFN). F.S. and A.S. are supported by ICSC-Centro Nazionale di Ricerca in High Performance Computing, Big Data and Quantum Computing, funded by European Union-NextGenerationEU  and by Italian Ministry of University and Research (MUR) projects FIS 0000155.

\appendix

\FloatBarrier

\bibliography{incbs}%

\end{document}